\documentclass[12pt]{article}
\usepackage{epsfig}
\usepackage{graphics}
\usepackage{color}
\usepackage{subfigure}
\usepackage{graphicx}
\usepackage{url}
\usepackage{multirow}
\usepackage{rotating}
\usepackage{longtable}
\usepackage{colortbl}
\renewenvironment{abstract}{%
    \setlength{\parindent}{0in}%
    \setlength{\parskip}{0in}%
    \bfseries%
    }{\par\vspace{12pt}}

\newenvironment{affiliations}{%
    \setcounter{enumi}{1}%
    \setlength{\parindent}{0in}%
    \slshape\sloppy%
    \begin{list}{\upshape$^{\arabic{enumi}}$}{%
        \usecounter{enumi}%
        \setlength{\leftmargin}{0in}%
        \setlength{\topsep}{0in}%
        \setlength{\labelsep}{0in}%
        \setlength{\labelwidth}{0in}%
        \setlength{\listparindent}{0in}%
        \setlength{\itemsep}{0ex}%
        \setlength{\parsep}{0in}%
        }
    }{\end{list}\par\vspace{12pt}}

\title{Information Measure for Long-Range Correlated Sequences:\\the Case of the 24 Human Chromosomes.}

\author{A. Carbone$^{1,2,3,*}$}

\begin{document}

\maketitle

\begin{affiliations}
 \item Politecnico di Torino, Italy
 \item ISC-CNR, Unit\`a Universit\`a `La Sapienza' di Roma, Italy
 \item ETH Zurich, Switzerland
\end{affiliations}

\begin{abstract}
A new approach to estimate the Shannon entropy of a long-range correlated sequence is proposed.
The entropy is written as the sum of two terms corresponding respectively to power-law  (\emph{ordered}) and exponentially (\emph{disordered}) distributed blocks (clusters). The approach is illustrated on  the 24 human  chromosome sequences by taking the nucleotide composition as the relevant information to be encoded/decoded.  Interestingly, the nucleotide composition of  the \emph{ordered} clusters is found, on the average, comparable to the one of the whole analyzed sequence, while that of the \emph{disordered} clusters fluctuates. From the information theory standpoint,  this means that the  power-law correlated clusters  carry the same information of the whole analysed sequence.   Furthermore, the fluctuations of the nucleotide composition of the disordered clusters are linked to relevant biological properties, such as segmental duplications and gene density.
\end{abstract}
Complex systems are probed by observing a relevant
quantity over a certain temporal or spatial range,
yielding long-range correlated
 sequences or arrays,
with the remarkable feature of  displaying `ordered' patterns, which emerge from the
seemingly random structure.  The degree of `order'  is intrinsically linked to the information embedded in the patterns, whose extraction and quantification might add clues to many complex phenomena \cite{Scheffer,Crutchfield,Wang,Grassberger,Steur,Shalizi,Carbone1,Carbone2,Carbone3,Carbone4,Shao}.
\par
In this work, an information measure for long-range correlated sequences, worked out from a partition of the sequence into \emph{clusters} according to the method proposed in \cite{Carbone1,Carbone2}, is put forward. The clusters are characterized by their length $\ell$, duration $\tau$ and area $\mathcal{A}$, obeying power-law probability distributions, with a cross-over to an exponential decay at large size.
The probability distribution function of the lengths  is considered to estimate the Shannon
(block) entropy $S(\ell)$ of the clusters. The entropy can be written as the sum of three terms,  respectively constant, logarithmic and linear function of the cluster length. The clusters with dominant logarithmic term of the entropy are power-law correlated and correspond to `ordered' structures, while those with dominant linear term are exponentially distributed and correspond to `disordered' structures.
The information measure is illustrated by analyzing the 24 nucleotide sequences
 of the human chromosomes.
Each sequence is first mapped
to a fractional
Brownian walk (the so-called DNA walk). Then, the probability distribution function $P(\ell)$  and the entropy $S(\ell)$ of the DNA clusters are estimated by adopting the proposed approach.
\par
It is worth recalling  that the investigation of the block entropy of a signal was originally motivated by cryptography. Claude Shannon attempt was aimed at encoding information in ways that still allowed recovery  by the receiver, the main question to be answered being: `How the signal can be compressed in elementary messages which still contain the relevant information to be communicated?'.
The approach proposed in this work represents a possible answer. Furthermore, this question recalls  the  concept of Kolmogorov complexity $KC(\ell)$  which quantifies the interplay of randomness/determinism of the strings output of a computational program. The Kolmogorov complexity  is quantified in terms of the minimal length of the program that can still generate a random string. It can be demonstrated that the length of the program, which is defined case-by-case in the specific computational framework, is comparable to the length of the string plus a constant, and varies as the logarithm of the length of the string itself.
\par
 From the information theory standpoint, the present work shows that by taking the nucleotide composition of the whole sequence as the relevant information to be transmitted from the source to the receiver, the whole sequence is encoded in blocks (clusters), which are able to transmit the same information of the whole sequence if they are power-law correlated.
Specifically, it is shown that the power-law correlated clusters are
characterized by a nucleotide content, purine-pyrimidine pairs (GC)\% and (AT)\%,
on the average equal to the value of the whole chromosome sequence under analysis.
Conversely, the exponentially correlated clusters are characterized by
a percentage of purine-pyrimidine pairs exhibiting fluctuations around the value taken by the whole sequence.  Interestingly, the standard deviationof the cluster composition fluctuations for each of the 24 chromosomes is correlated to biologically relevant properties, such as duplication frequency and gene density.
It is worthy of remark that the nucleotide composition  is taken as a case study  for the illustration of the implementation and meaning of the proposed entropy measure, but it is not the only biologically relevant information carried by a DNA sequence.
%
%
%
\par
\subsection*{Results}
The  entropy of a sequence, coded in blocks, has been extensively studied since its introduction by Shannon (see e.g. \cite{Crutchfield,Wang,Grassberger,Steur}, and Refs. therein). The practical application of the Shannon  entropy concept requires a symbolic representation of the data, obtained by a suitable partition transforming the continuous phase-space into disjoint sets. As discussed in \cite{Steur}, the construction of the optimal partition is not a trivial task, being crucial to effectively discriminate between randomness/determinism of the encoded/decoded data. The method commonly adopted for  partitioning a sequence and estimating its entropy is based on a uniform division in blocks having equal length $\ell$. Then the entropy is estimated over subsequent partition corresponding to different blocks lengths $\ell$. The novelty of the present work resides in the method used for partitioning the sequence which directly yields   power-law or exponential distributed blocks (clusters). This is a major advantage, as it allows one to straightforwardly separate  the set of inherently correlated/uncorrelated blocks along the sequence.
\par
\begin{figure}
\includegraphics[width=7cm,height=5cm,angle=0]{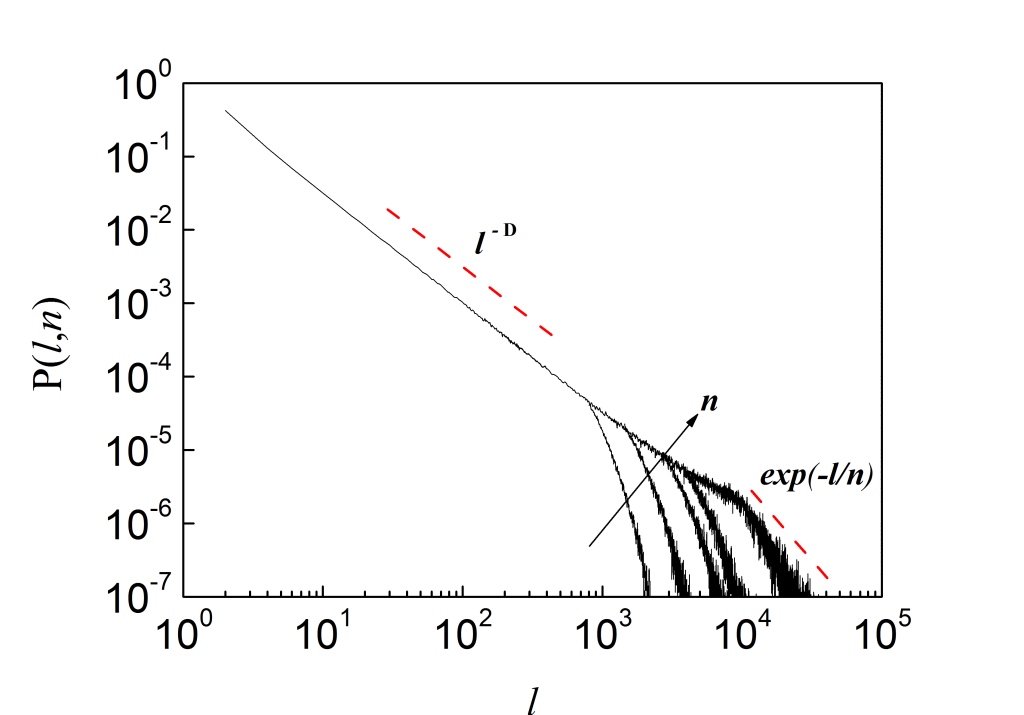}
\centering \caption {\label{figure1}  Cluster Length Probability Distribution. Probability distribution
function $P(\ell,n)$ of cluster lengths for a sequence with $H\approx 0.6$ and $L=2^{20}$. The moving average windows
are $n=500$, $n=1000$, $n=2000$, $n=3000$ and $n=10000$ (from left
to right). As $n$ increases, $P(\ell,n)$ becomes broader. The
slope of the distribution becomes steeper for $\ell>n$,
corresponding to the onset of finite-size effects and exponentially
decaying correlation.}
\end{figure}

A random sequence  $y(x)$  can be partitioned in elementary
clusters  by the intersection with the moving average $\widetilde{y}_n(x)$
where $n$ is the size of the moving window. The clusters correspond to the regions bounded
by $y(x)$ and $\widetilde{y}_n(x)$ between two subsequent
crossings points ${x_c(i)}$ and ${x_c(i+1)}$ \cite{Carbone1}.
The
intersection between  $y(x)$ and  $\widetilde{y}_n(x)$ produces a {\em
generating} partition, yielding different sequences of clusters
for different values of $n$.
 The
probability distribution function  $P(\ell,n)$ of the
lengths $\ell$ for each $n$ can be obtained by counting the clusters
${\mathcal N}(\ell_1,n),{\mathcal N}(\ell_2,n), ..., {\mathcal
N}(\ell_i,n)$ respectively with length $\ell_1, \ell_2,...,
\ell_i$. By doing so, one obtains \cite{Carbone1}:
\begin{equation}
\label{Pl} P(\ell,n)\sim\ell^{-D} {\mathcal F}\left({\ell},{n}\right)\sim \mu\left({\ell},{n}\right)^{-1}
\hspace{5pt},
\end{equation}
where  $D=2-H$ and $H$  indicate respectively the fractal dimension  and  the Hurst exponent of the sequence.  The exponent $H$ is widely used for quantifying long-range correlations  (power-law decaying) as opposed to short-range (exponentially decaying)  correlations   in many complex  systems. The Hurst exponent has been estimated for the 24 chromosome sequences, as reported in the 3$^{rd}$ of Table 1.  The occurrence of long-range correlations means that the nucleotides are organized along the sequences in similar way, a fact that can be defined as \emph{compositional self-similarity} of the chromosomes.
The function ${\mathcal F}\left({\ell},{n}\right)$ in equation \ref{Pl} can be taken of the form:
\begin{equation}
\label{e.4x} {\mathcal F}(\ell,n) \equiv \exp({-\ell}/{n})\hspace{5pt}.
\end{equation}
${\mathcal F}\left({\ell},{n}\right)$ accounts  for the drop-off of  $P(\ell,n)$
 due to finiteness of $n$ when $\ell\gg n$. The quantity ${\mu}(\ell,n)\sim \ell^D \exp({\ell/n})$ is proportional to the size of the subsets spanned by the random walkers which ranges  from a line proportional to $\ell$ for
$H=1$  to a  square  proportional to $\ell^2$ for $H=0$ for $n >\ell$.
The probability distribution function $P(\ell,n)$ is shown in Fig.~\ref{figure1} for a wide
range of $n$ values, estimated for a long range correlated series with Hurst exponent $H \approx 0.6$.
 For $n \to 1$, the lengths $\ell$ of the
elementary clusters are centered around a single value. When
$n$ increases, a broader range of lengths is obtained and,
consequently, $P(\ell,n)$ spreads over all values.
\par
The Shannon entropy is defined as \cite{Crutchfield,Wang,Grassberger,Steur}:
\begin{equation}
\label{lentropy} {S}(\ell,n)\equiv
-\sum P(\ell,n)\log P(\ell,n)\hspace{5pt},
\end{equation}
where the sum is performed over the number of elementary clusters  with length $\ell$  obtained by the intersection with the moving average for each $n$. This number ranges from $1$ to ${{\mu} (\ell,n)}^{-1}$ depending on how many clusters are generated by the intersection with the moving average. The value $1$ is obtained when only one cluster with length $\ell$ is found in the partition.  As already noted, the standard method for partitioning a sequence and estimating its entropy is by splitting the sequence into a set of disjoint blocks with equal length $\ell$.  Conversely, in the present work,  the intersections of the sequence with the moving average generate a set of disjoint blocks with a broad distribution of lengths $\ell$ corresponding respectively to power-law or exponential correlation. This particular partition retains the determinism/randomness of the blocks by simply varying  $n$, an aspect intimately related to the Kolmogorov complexity concept.
\par
By using equations~(\ref{Pl}) and (\ref{lentropy}), the cluster entropy writes:
\begin{equation}
\label{lentropy1} S(\ell,n)=S_0+\log\ell^D-\log{\mathcal
F}(\ell,n),
\end{equation}
which, after taking into account equation~(\ref{e.4x}), becomes:
\begin{equation}\label{lentropy2}
S(\ell,n)=S_0+\log\ell^D+{\ell\over n}\hspace{5pt},
\end{equation}
where $S_0$ is a constant, $\log\ell^D$ is related to the term $\ell^{-D}$ and $\ell/ n$ is related to the term ${\mathcal F}(\ell,n)$.
\par
To clarify the meaning of the terms appearing in equation~(\ref{lentropy2}), it is worthy of remarking that for \emph{isolated
systems}, the entropy increase $dS$ is related to the  irreversible processes
spontaneously occurring within the system. The entropy tends to a constant value as a stationary
state is asymptotically reached ($dS\geq 0$). For \emph{open systems} interacting with their
environment, the increase is given by a term $dS_{int}$,
 due to the  irreversible processes
spontaneously occurring within the system, and a term $dS_{ext}$
 due to the irreversible processes arising through the external interactions.
The term $\log\ell^D$ in equation~(\ref{lentropy2}) should be interpreted as the intrinsic entropy $S_{int}$. It
is indeed independent of $n$, i.e. it is independent of the method  used for  partitioning the sequence,
which plays here the role of the external interaction. The logarithmic term is of the form of
a Boltzmann entropy $S=\log\Omega$, where $\Omega$ is the
maximum volume occupied by the isolated system. The quantity $\ell^D$ corresponds  to the
volume occupied by the random walker. Whenever $\ell$
could reach the maximum size $L$ of the sequence, the second term
on the right side would write $\log {L^{D}}$.
The term $\ell/n$ in equation~(\ref{lentropy2})
represents  the excess entropy $S_{ext}$ introduced by the partition process. It comes into play when the sequence is partitioned in clusters and depends on $n$.
 \par
\begin{figure}
\includegraphics[width=7.0cm,height=5cm,angle=0]{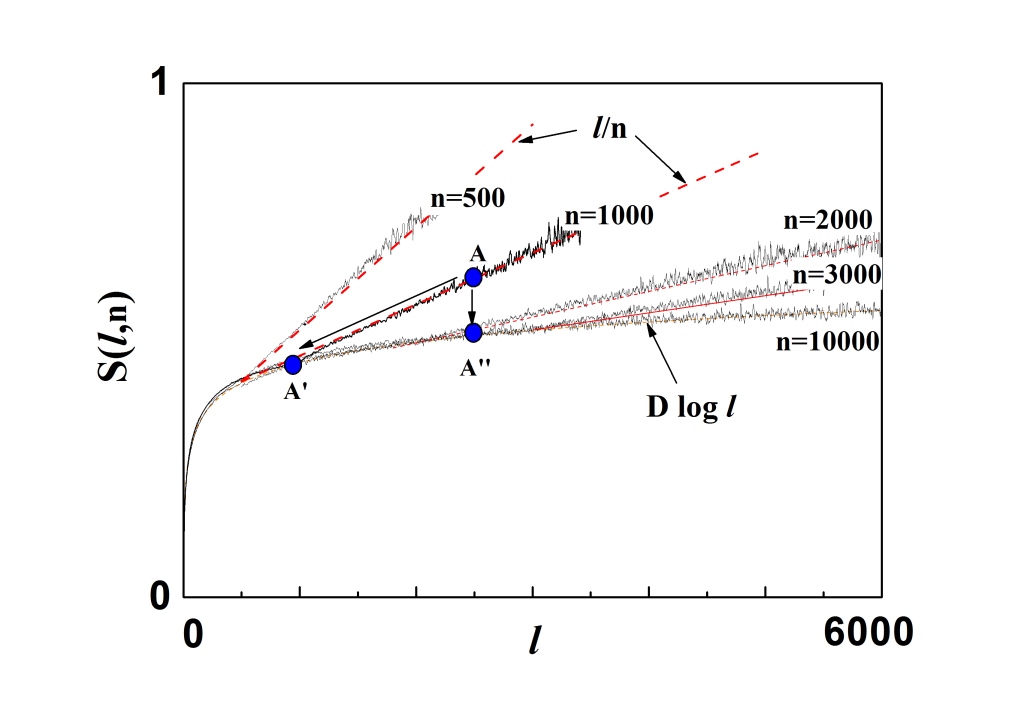}
\centering \caption {\label{figure2} Cluster Entropy. Entropy $S(\ell,
n)$  of the clusters corresponding to the probability distribution function $P(\ell,n)$
plotted in Fig.~\ref {figure1}. For small values of $\ell$, the curves
increase logarithmically as $\log \ell^D$ and are $n$-invariant, while they vary as a linear function for larger values of $\ell$, as expected according to equation~(\ref{lentropy2}).}
\end{figure}

Fig.~\ref{figure2} shows the entropy ${S}(\ell,n)$ evaluated by
using the probability distribution $P(\ell,n)$ plotted in Fig.~\ref{figure1}.
 One can note that ${S}(\ell,n)$
increases logarithmically as $\log \ell^D$ and is $n$-invariant for small values of $\ell$, while it increases as a linear function  at larger $\ell$, as expected according to equation~(\ref{lentropy2}). Clusters with lengths $\ell$ larger than $n$
are not power-law correlated, due to the finite-size
effects introduced by the window $n$. Hence, they are characterized
by a value of the entropy exceeding the curve $\log \ell^D$, which corresponds to power-law correlated clusters.
 It is worthy to remark that clusters with a given length $\ell$
  can be generated by different values of the window $n$.  For example, clusters with $\ell=2500$ have entropies corresponding to the point $A$ (for $n=1000$) or $A''$ (for $n=3000$ and $n=10000$) as shown in Fig.~\ref{figure2}.  One can observe that  $A''$  corresponds to power-law correlated
(ordered) clusters, since $A''$ lies  on the curve
$\log\ell^D$. Conversely, the point $A$
 does not correspond to power-law correlated clusters, since  $A$ lies on the curve $\ell/n$ which originates from the term $\mathcal F(\ell,n)$.
In other words,  clusters
with lengths shorter than $n$ are ordered (long-range correlated), whereas clusters
with lengths larger than $n$ are disordered (exponentially correlated).
\par
To gain further insight in the meaning of the terms appearing in
equation~(\ref{lentropy2}), the
\emph{source entropy rate} $s$ is calculated for the entropy ${S}(\ell,n)$. The
source entropy rate is a measure of the \emph{excess randomness} and increases as the block coding process becomes noisier. By using the definition and equation~(\ref{lentropy2}), the
source entropy rate writes:
\begin{equation}
\label{rate0}
    s\equiv\lim_{\ell\rightarrow\infty} {S(\ell,n)\over \ell} = \frac{1}{n} \hspace{5pt}.
\end{equation}
The excess randomness of the clusters is found to be inversely proportional to $n$ and, thus, becomes negligible in the limit of $n \rightarrow \infty$. This clearly occurs  in the curves of Fig.~\ref{figure2},
where one can note that higher entropy rates correspond to steeper slopes of the linear term $\ell/n$ (smaller $n$ values).


\subsection*{Discussion}
In this section, the  information measure is implemented on the 24 human chromosomes, mapped to fractional Brownian walks (mapping details are described in Method).
The nucleotide composition of the DNA sequence is taken as the relevant information quantity to be encoded from the source and decoded from the receiver.
\par
It is well-established that the two strands of DNA are held together by hydrogen bonds between complementary bases: two bonds for the AT
pair and three bonds for the GC pair, which is therefore stronger.
The existence of GC-rich and GC-poor segments may play different roles in biological
processes as duplication, segmentation, unzipping \cite{Lander,Bailey,Deloukas}.
\par
Nonuniformity  of nucleotides composition within genomes  was revealed
several decades ago by thermal melting and gradient centrifugation.
On the basis of findings concerning buoyant densities of
melted DNA fragments, a theory for
the structure of genomes of warm-blooded vertebrates  known as the
\emph{isochores theory} was put forward \cite{Lee,Bernardi,Costantini,Clay}.  Isochores were defined as
 genomic segments that are fairly homogeneous in their guanine
and cytosine (GC) composition.
\par
Though it is widely accepted
that the human genome contains large regions of distinctive GC
content, the availability of fully sequenced DNA or RNA
molecules allows one to accurately investigate the
local structure  by statistical methods.
The development of  efficient algorithms achieving deep and accurate description of the complex genomic architecture is thus a timely endeavour \cite{Cohen,Versteeg,Emanuel,Vaillant,Li,Salerno,Peng,Roman,Akhter}.
\par
\begin{figure}
\includegraphics[width=18.0cm,height=12cm,angle=0]{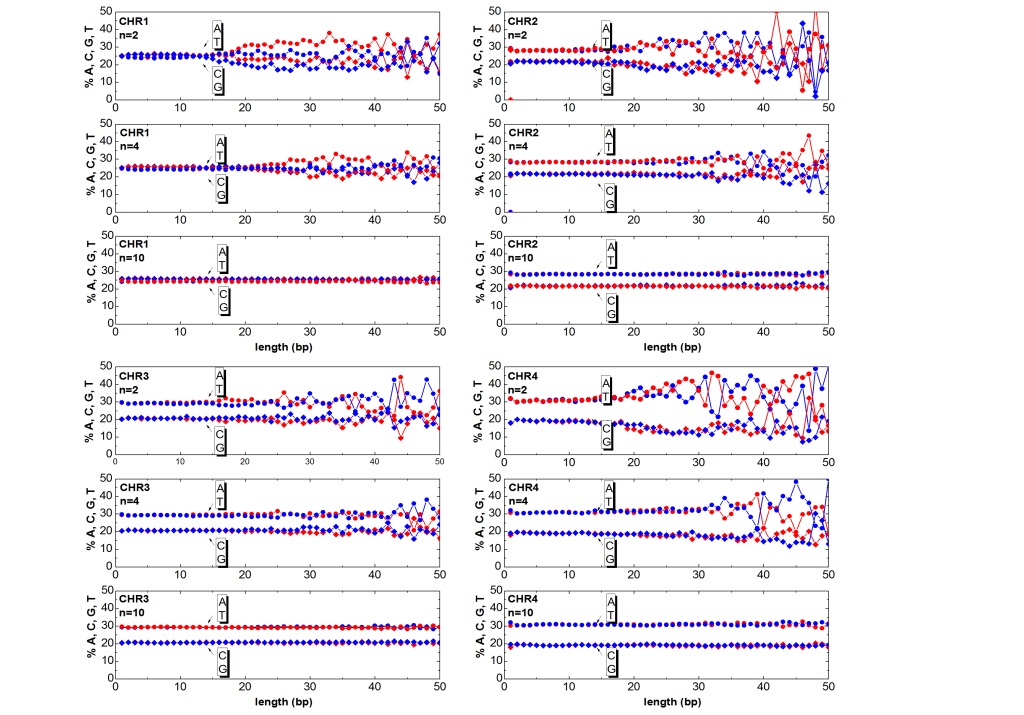}
\centering \caption {\label{DNA1}  Cluster Composition.  Base composition (\% of A (blue) T (red) C (blue) G (red) nucleotides) of the clusters in the human chromosomes 1, 2, 3, 4. For each chromosome, the plots refer to  windows $n=2$, $n=4$, $n=10$.  Data refers to the first 10Mbases of each chromosome. See Tables S1-S6 of Supplementary Information for further estimates.}
\end{figure}
\begin{figure}
\includegraphics[width=18.0cm,height=12cm,angle=0]{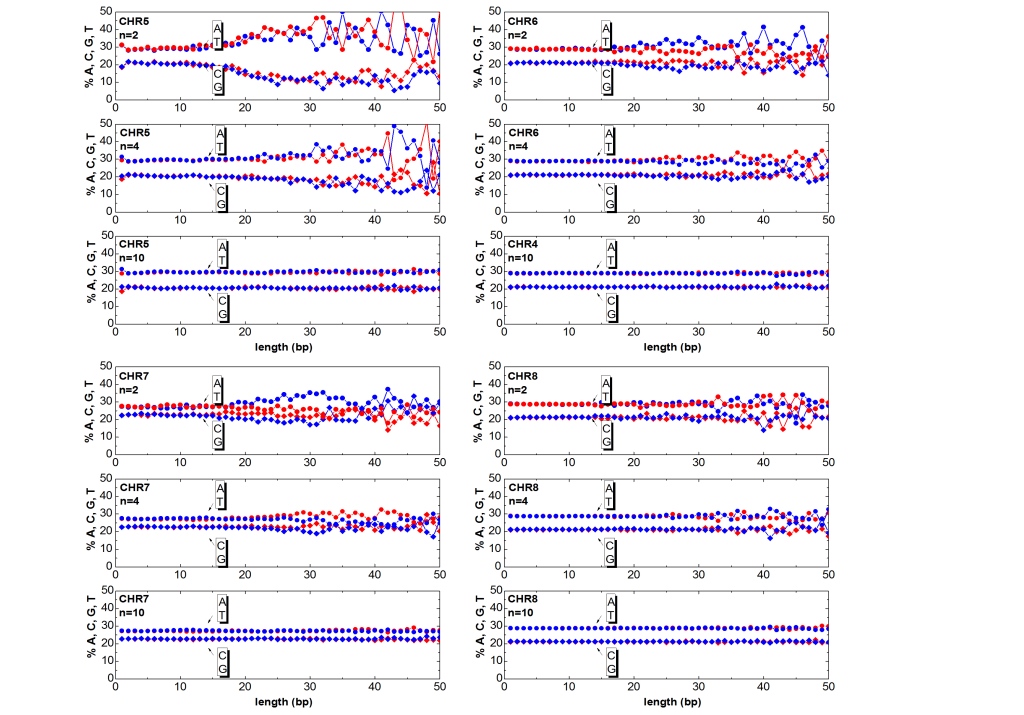}
\centering \caption {\label{DNA2}    Same as Fig.~\ref{DNA1} but for the chromosomes 5, 6, 7, 8.}
\end{figure}
\begin{figure}
\includegraphics[width=18.0cm,height=12cm,angle=0]{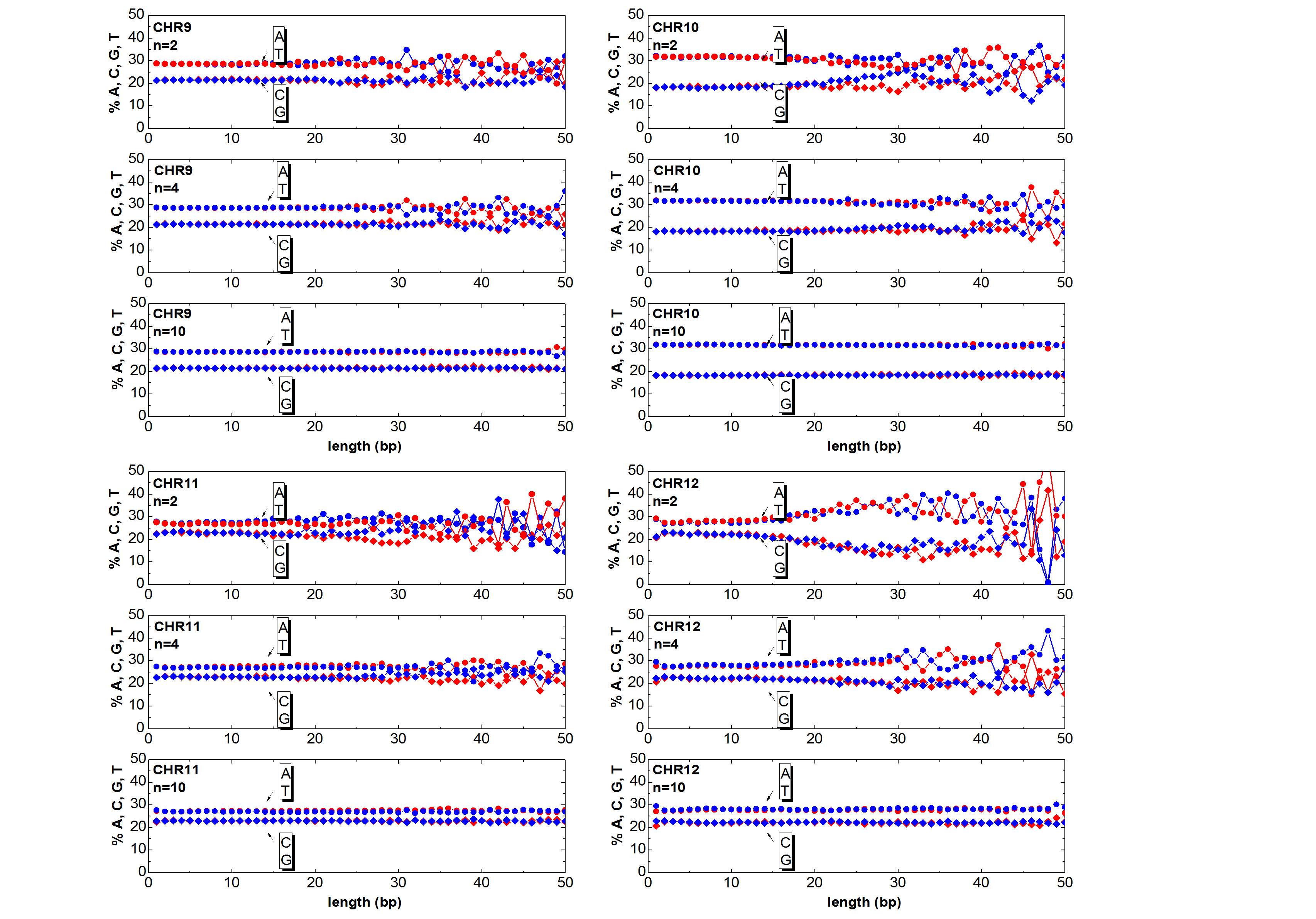}
\centering \caption {\label{DNA3}    Same as Fig.~\ref{DNA1} but for the chromosomes 9, 10, 11, 12.}
\end{figure}
\begin{figure}
\includegraphics[width=18.0cm,height=12cm,angle=0]{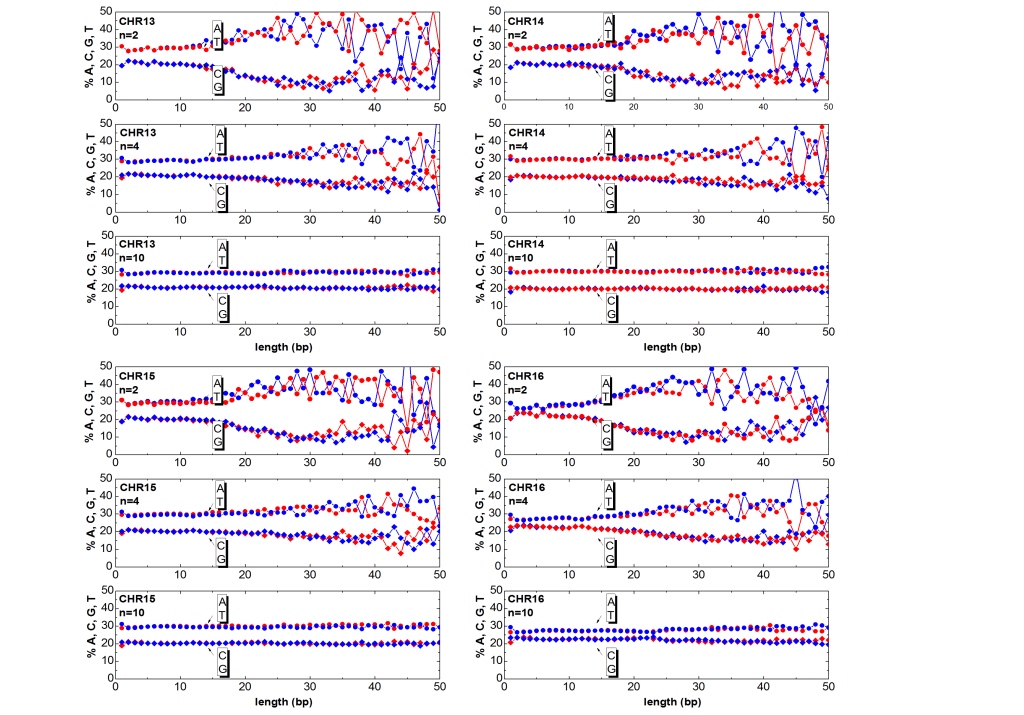}
\centering \caption {\label{DNA4}   Same as Fig.~\ref{DNA1} but for the chromosomes 13, 14, 15, 16.}
\end{figure}
\begin{figure}
\includegraphics[width=18.0cm,height=12cm,angle=0]{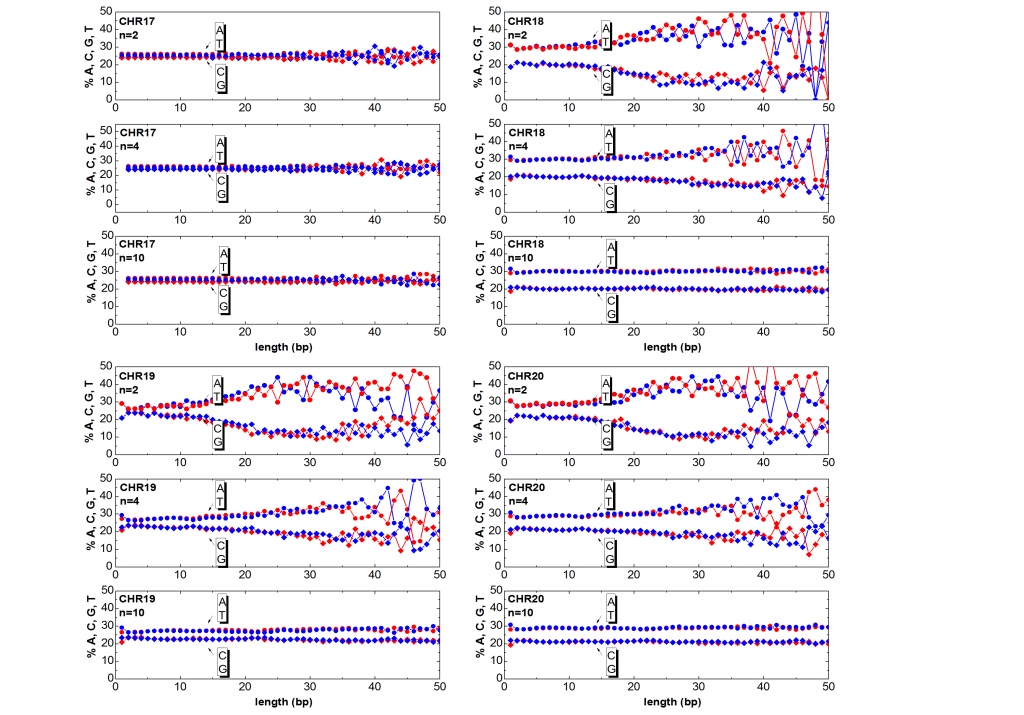}
\centering \caption {\label{DNA5}    Same as Fig.~\ref{DNA1} but for the chromosomes 17, 18, 19, 20.}
\end{figure}
\begin{figure}
\includegraphics[width=18.0cm,height=12cm,angle=0]{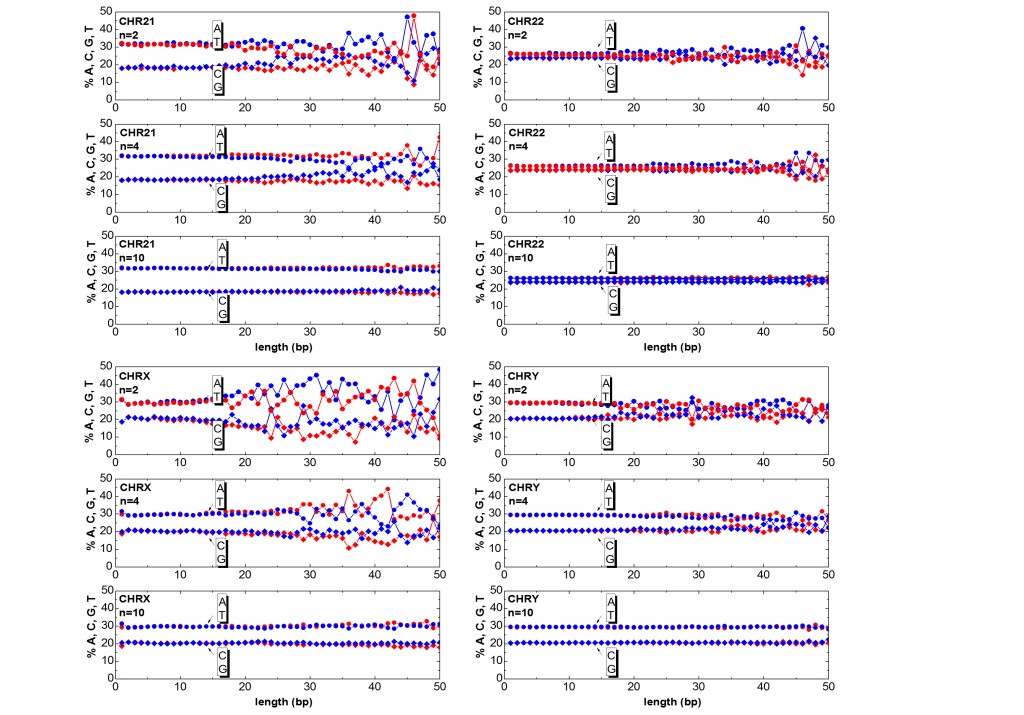}
\centering \caption {\label{DNA6}   Same as Fig.~\ref{DNA1} but for the chromosomes 21, 22, X, Y.}
\end{figure}

The chromosomes can be mapped to numeric sequences according to different approaches. In this work, first the DNA is mapped (as detailed in the section Method) to a random walk,  then the clusters are generated as described in the previous section. Once having generated the clusters, one can answer the question `How much of the relevant information is still contained in the clusters?'. The answer to this question is obtained by counting the ATGC basis for each cluster and plotting the percentage as a function of the cluster length.
In Figs. \ref{DNA1}-\ref{DNA6}, the nucleotide
compositions are plotted as a function of the cluster length $\ell$ for  $n=2$, $n=4$ and $n=10$. The range of
$n$ values used in this work  varied  from 2 to 10.000.
One can observe that the nucleotides count is roughly constant for clusters having length comparable
or shorter than $n$. This means that \emph{ordered} DNA clusters with constant nucleotide composition are
found, when the entropy varies as a logarithm of $\ell$. For
cluster lengths $\ell$ larger than $n$, the power-law correlation breaks down
with the onset of exponentially correlated clusters (`disordered' clusters).
An even more interesting result is that the amplitude of the fluctuations is not constant as it takes a characteristic value for each chromosome. One can note from the data plotted in  Figs. \ref{DNA1}-\ref{DNA6} that the fluctuations of the cluster composition is very small for example in chromosomes 8, 9, 17, Y. Conversely, they are quite large for chromosomes 14, 15, X.
 It should be remarked that Figs. \ref{DNA1}-\ref{DNA6} show the  nucleotide composition of the ordered-disordered clusters. These plots are related to the entropy of the blocks if one bears in mind the original aim of the Shannon work. The estimate of the block entropy was originally motivated by the attempt at decoding information in ways that still allow   recovery of the relevant information  by the receiver. In other words, the main question raised by Claude Shannon is: ``How the signal can be compressed in elementary messages (blocks) which still contain the relevant information to be communicated?".
The approach proposed in this work  answers this question. The DNA sequence is encoded in short messages (clusters) able to transmit the same information of the whole sequence (from where they were cut out) only if they are power-law correlated. In this manuscript, the information considered relevant to the receiver is the nucleotide composition, which, of course,  is not the only choice for the relevant information to be transmitted, as other characteristic features might be interesting as well.
It is also discussed to what extent nucleotide fluctuations, characterizing the exponentially correlated clusters of each chromosome, might be  linked to features relevant to biological processes. To this purpose, the standard deviation of the fluctuations has been calculated for the nucleotide composition ATGC of the clusters (values are reported in Table~\ref{tab:variance}). The correlation $\sigma_C$ with bilogical features characteristic  of each chromosome, such as  length,  gene density, inter-chromosomal duplications, intra-chromosomal duplications, local ATGC composition (data taken from Refs. \cite{Bailey,Deloukas}) have been considered. The correlation coefficients $\rho_C$ are shown in Table \ref{tab:correlation}. Negative correlations between $\sigma_C$ and intra-and inter-chromosomal duplications are found. Conversely, strong positive correlations are observed between $\sigma_C$ and AT-rich regions. These findings might point to the important result that the  cluster fluctuations are fingerprints of recent segmental duplications.
\subsection*{Methods}
\begin{figure}
\includegraphics[width=7.0cm,height=5cm,angle=0]{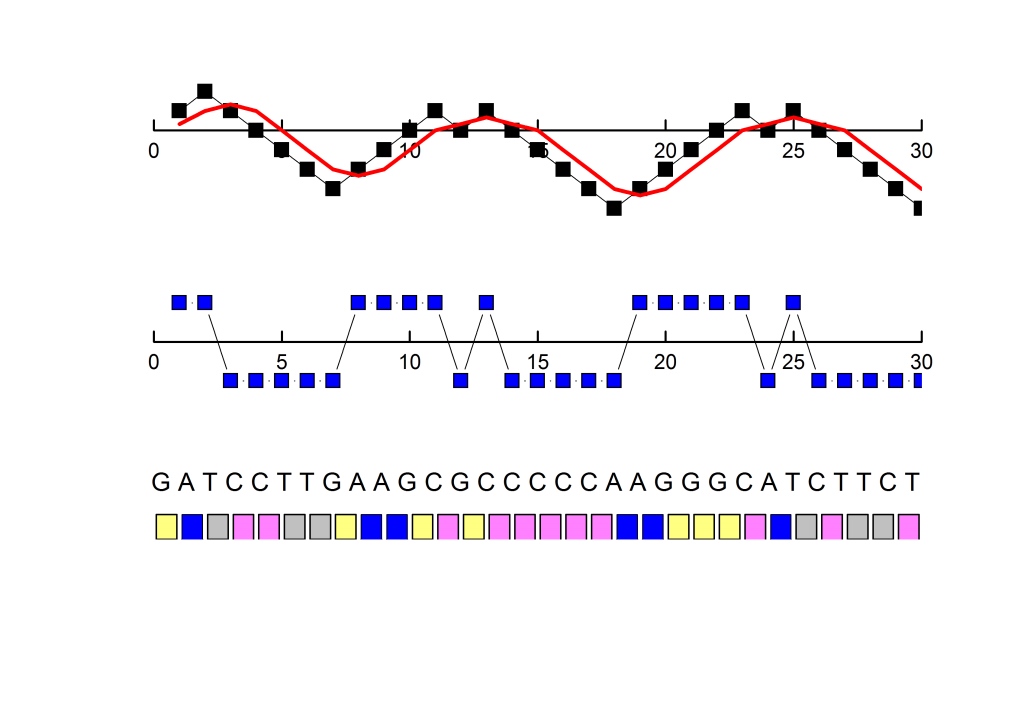}
\centering \caption {\label{figure3}   DNA Sequence Mapping Visualization. Bottom: scheme of the first 30 ATGC bases of the  sequence of the human chromosome 1. Middle:  the  sequence of +1 and -1 corresponding to the ATGC. Top: the DNA walk $y(x)$ obtained by summing the sequence of +1 and -1 (black squares) with the moving average $\widetilde{y}_n(x)$ with $n=3$ (red curve). }
\end{figure}

A DNA sequence is composed of four nucleotides: adenine (A), thymine (T), cytosine (C) and guanine (G). The first step of the analysis consists in the conversion of the four-letter genome alphabet  into a numerical format.
There are several ways of mapping a DNA sequence to a walk: one-dimensional up to 4 dimensional, real or complex representations. As  the proposed Shannon entropy measure applies to one-dimensional sequences, the present discussion is limited to  one-dimensional real representation of the four nucleotide bases.
The sequence of the nucleotide bases  is mapped according to the following rule:  if the base is a purine (A,G), the base is mapped to
$+1$, otherwise  if the base is a pyrimidine (C,T), the base is mapped to
$-1$ (Fig. \ref{figure3}). The sequence of $+1$ and $-1$  is summed and a random walk $y(x)$ (\emph{DNA walk}) is obtained. This coding rule is preferable, as it keeps the nonstationarity of the  series at a minimum. Large nonstationarity of the numerical series might be an issue when long-range correlation should be investigated. The average concentration of A and T are about 0.30, those of G and C are about 0.20. The concentration of purines (A+G) and pyrimidines (C+T) are very close to 0.50 along the sequence. Therefore, coding of  purines and pyrimidines to +1 and -1 guarantees a high degree of symmetry of the numerical  series.  Conversely, an asymmetric  coding rule would amplify the strong variations of the local density distribution of the bases along the sequences, giving rise to higher nonstationarity of the corresponding random walk.
\par
The  function $\widetilde{y}_n(x)$  is  calculated   for the \emph{DNA walk}  with different values of the window $n$.
The intersection between   $y(x)$ and  $\widetilde{y}_n(x)$ yields a set of clusters, which correspond to  the  segments between two adjacent
intersections  of $y(x)$ and $\widetilde{y}_n(x)$.
Since each cluster of the \emph{DNA walk} corresponds to  a cluster of ATGC  nucleotides,  the number of nucleotides can be counted  and plotted as a function of the length $\ell$ for each cluster.
In Figs. \ref{DNA1}-\ref{DNA6} the nucleotide composition of the clusters as a function of the length $\ell$  is shown for the 24 human chromosomes. The clusters have been cut out of  10$^6$ bases of each chromosome at once.
 To be statistically meaningful, there is a need to operate over subsequences having the same length (note  that the 24 human chromosomes have different lengths $L$, 2nd column of Table 1 ).  The method proposed here has been however implemented on several sequences with different lengths (varying from 10$^5$ to 10$^7$ have been considered in this study). This range takes into account that, on one hand,  a scaling law is sound when it is observed at least over three decades of a logarithmic scales, and the computational time and complexity on the other hand.  One can note that the average composition of the power-law correlated clusters  is comparable with the composition of the whole sequence of the analysed data. For example the nucleotide composition of the power-law correlated  clusters of the chromosome 1 should be confronted with the data reported in the column 8$^{th}$, 9$^{th}$, 10$^{th}$, 11$^{th}$ of Table 1 for the same chromosome, while  the standard deviation is reported in Table 2. The statistical robustness of the method has been checked by estimating the correlation coefficient $\rho_c$  of the variance and other biological parameters of the sequences (Table 3).
\par
One common problem  in data mining is the statistical validation of the model envisioned to describe data structures and patterns. The error is estimated on the entire sample set for small quantity of data.  For large data sets, more sophisticated cross-validation methods have been developed to quantify the performance of algorithms and models over disjoint subsets.  Depending upon the criterion used to split the data, the process of training and validation across disjoint sets is named \emph{random}, \emph{$k$-fold} or \emph{leave-one-out} \cite{Hastie}. In particular, the leave-one-out  is the degenerate case of the $k$-fold cross-validation, with only one disjoint subset ($k=1$) and is particularly useful for very sparse datasets with few samples, though its error might be larger than the error of the estimates themselves and computation time might be quite long. As the  analysed dataset (the 24 genomic sequences) is large enough, the random and $k$-fold cross validation can be used with the advantage of higher accuracy and velocity of the estimates. In Supplementary Tables S1-S6, the average values and variances of the nucleotide contents obtained over three disjoint data sets are reported for the 24 chromosomes. For each subset, when the parameter $n$ is varied,  clusters of any lengths are generated in random position of the sequence allowing to estimate the average composition and the statistical errors at different position along the sequence. For each set the standard deviations are also reported in the Supplementary Tables S1-S6.
\par
 Finally, we note that the Hurst exponent for the 24 chromosomes is reported in the  3$^{rd}$ column of  Table 1. As one can see the value of the exponent $H$ is higher than 0.5, implying that a positive correlation (persistence) exist among the nucleotides. The values of the Hurst exponents have been obtained by using the method described in Refs. \cite{Carbone1,Carbone2,Carbone3}.
 \par
The sequences used in this analysis
were retrieved from the NCBI ftp server (\url{ftp://ftp.ncbi.nlm.nih.gov/genomes/H_sapiens/}).
%
%
%
%
\subsection*{References}

\subsection*{Additional Information}
The author declares to have no competing financial interests.\\
Correspondence and requests for materials should be addressed to anna.carbone@polito.it

\begin{table*}
\begin{tabular}{c|cccccc|cccc}
 \hline \hline
  CHR & $L$ & $H$ & A [\%] & C[\%] & G[\%] & T[\%] & A[\%] & C[\%] & G[\%] & T[\%]\\
  \hline
  1 & 226217758 & 0.64 & 29.09 & 20.87 & 20.87 & 29.14 & 26.52  & 25.79  & 25.58  & 25.15  \\
   \hline
  2 & 237900011 & 0.66 & 29.84 & 20.11 & 20.13 & 29.90 & 28.50 & 24.51 & 22.34   & 29.81  \\
   \hline
  3 &  195304882 & 0.66 & 30.14 & 19.84 & 19.84 & 30.16   & 28.46  & 21.77 & 21.50  & 28.65\\
  \hline
  4 &  187941502 & 0.66 & 30.87 & 19.11 & 19.12 & 30.88  & 34.47  &  19.80  & 22.86  & 30.28  \\
  \hline
  5 & 177847050 & 0.66  & 30.20 & 19.74 & 19.77 & 30.27   & 29.97  & 24.86  & 19.70  & 36.43  \\
   \hline
  6 &  169100547 & 0.65 & 30.18 & 19.80 & 19.81 & 30.19   &29.97 & 21.23  & 21.73 & 28.27 \\
   \hline
     7 & 155403473 & 0.66 & 29.60 & 20.38 & 20.36 & 29.63  & 28.85 & 21.79   & 27.09  & 22.93 \\
     \hline
  8 &  143332430 & 0.65 & 29.90 & 20.06 & 20.06 & 29.86   &29.51  & 20.61  & 20.85   & 29.03\\
  \hline
  9 &  120994158 & 0.67 & 29.35 & 20.65 & 20.64 & 29.33  & 27.91  & 21.83 & 21.38 & 28.76 \\
  \hline
  10 &  131739836 & 0.65 & 29.19 & 20.79 & 20.78 & 29.22 &31.15  &  19.94  & 19.47  & 29.44 \\
  \hline
  11 &  131247160& 0.68 & 29.20 & 20.77 & 20.79 & 29.21   & 28.97  & 22.23    & 25.08  & 26.85  \\
  \hline
  12 &  130304143& 0.67 & 29.59 & 20.40 & 20.39 & 29.60  & 30.66 & 22.85 &  24.19  & 29.62  \\
  \hline
  13 &  95747346& 0.66 & 30.69 & 19.26 & 19.26 & 30.77  & 33.94 &  20.95& 21.82 & 33.88  \\
  \hline
   14 & 88290585& 0.67 & 29.44 & 20.41 & 20.46 & 29.67  & 33.94 &  20.95& 21.82 & 33.88  \\
   \hline
  15 & 81927784 & 0.66& 28.89 & 21.13  & 21.10 & 28.86  & 32.64  & 21.69  & 20.74  & 33.00  \\
   \hline
    16 & 78990748& 0.67 & 27.53 & 22.35 & 22.44 & 27.66  & 29.17  & 24.30   &22.89  & 31.49 \\
     \hline
  17 &  79620483 & 0.65& 27.17 & 22.81 & 22.76 & 27.22  & 25.36 & 24.80  & 24.73 & 24.87 \\
   \hline
   18 &  74660927& 0.67 & 30.09 & 19.87 & 19.90 & 30.12  & 25.36 & 24.80  & 24.73 & 24.87\\
   \hline
   19 &  56038018& 0.66 & 25.79 & 24.14 & 24.20 & 25.86& 32.65  & 21.61 & 23.15  & 30.64  \\
   \hline
   20 &  59505758& 0.66 & 27.76 & 22.02 & 22.09 & 28.10  & 29.01 & 24.09  & 19.02   & 36.45 \\
   \hline
  21 & 35452914& 0.65 & 29.68  & 20.39 & 20.44  & 29.46 & 32.27  & 19.25 & 21.18    & 27.29 \\
   \hline
  22 & 35059666& 0.65 & 26.08  & 23.98  & 23.95  & 25.96  & 28.30    &22.93    & 24.63  & 24.92 \\
\hline X & 152580014 & 0.65 & 30.20 & 19.73 & 19.76 &30.26   & 32.88  & 20.86  & 25.39  & 28.01\\
\hline
  Y & 25654723& 0.72 & 29.88& 19.87 & 20.08 & 30.14  & 27.45  & 22.05  & 24.21  & 26.49\\
  \hline \hline
\end{tabular}
\caption{\label{tab:table1} Nucleotide Composition. Length $L$ (2$^{nd}$ column), Hurst exponent $H$ (3$^{rd}$ column), base composition (\% of ATCG, 4$^{th}$-7$^{th}$ columns) of the 24 chromosome
whole sequences.  Average nucleotide composition (\% of the ATCG, 8$^{th}$-11$^{th}$ columns) of the clusters, estimated according to the proposed method with $n=4$ over the first 10MBases
of the 24 chromosome sequences. In particular, the data in the  8$^{th}$-11$^{th}$ columns correspond to the plots shown in the  middle panels of Figs.~3-8 for each chromosome. In Tables S1-S6 of Supplementary Information, further results, estimated over different  data sets with different values of $n$, are reported.
}
\end{table*}

\begin{table*}

\begin{tabular}{c|cccc}
 \hline \hline
  CHR  & $\sigma_C$ [A]  & $\sigma_C$ [C]  & $\sigma_C$ [G]  & $\sigma_C$ [T] \\
  \hline
1&  11.01  & 10.81  & 10.06  & 9.68  \\  \hline
2& 14.19  & 12.90   & 12.43 & 14.30 \\  \hline
3&   10.05  & 9.43  &  8.29   & 8.52 \\  \hline
4&  16.56 & 12.91  & 13.87  & 17.79 \\  \hline
5&  19.75  & 14.32 & 12.76 & 16.69 \\  \hline
6&  9.07   & 6.33  & 7.42 &  8.71  \\   \hline
7&  8.58  & 8.32  & 11.14  & 10.21 \\   \hline
8&  4.89  &  3.88  & 4.33  &  4.91 \\   \hline
9&  6.49  &  4.97   & 4.36 &  5.23 \\  \hline
10&  10.84  &  9.04  &  7.52   & 9.38 \\  \hline
11&   9.12   & 8.83   & 11.69 & 10.87 \\ \hline
12&   17.09  & 14.86  & 14.13 & 15.63\\  \hline
13&   17.12  & 12.01  & 13.78  & 18.11 \\  \hline
14&   19.53  & 13.06  & 13.43  & 19.26 \\ \hline
15&   16.55 & 14.08   & 12.54  & 16.61\\  \hline
16&   16.31  & 15.60   & 15.77   & 15.69 \\ \hline
17&    5.06  &  5.17   & 4.95   & 5.67 \\ \hline
18&   20.02  & 13.98   & 14.10  & 18.47 \\ \hline
19&   17.90  & 14.88   & 13.99  & 17.10 \\ \hline
20&  17.31  & 13.86   & 14.33  & 18.70 \\  \hline
21&   10.84  &  7.69   & 8.50  & 10.81 \\  \hline
22&   8.40   & 5.81  &  5.53  &  8.48 \\ \hline
X &   18.06 & 14.09 & 14.87  &19.06 \\  \hline
Y &   6.19    & 6.66 &  7.08  &  7.47 \\ \hline \hline
\end{tabular}
\caption{\label{tab:variance}  Standard deviation  of the cluster nucleotide composition.  Standard deviations refer to the average values   (\% of the ATCG, 8$^{th}$-11$^{th}$ columns), estimated according to the proposed method with $n=4$ over the first 10MBases
of the 24 chromosome sequences.  Standard deviationscan be appreciated in the  middle panel plots of Figs.~3-8 for each chromosome. In the Tables S1-S6 of Supplementary Information, further values over different chromosome sets and with different values of $n$ are reported.}
\end{table*}

\begin{table*}

\begin{tabular}{cccc}
\hline \hline
  & $\rho_C$ [M$_1$] & $\rho_C$ [M$_2$] & $\rho_C$ [M$_3$]  \\
  \hline
Length & -0.194 & -0.552 & -0.582  \\  \hline
Gene density  & -0.178 & -0.076 & -0.107 \\  \hline
Inter-chromosomal duplications & -0.330 & -0.242 & -0.165 \\  \hline
Intra-chromosomal duplications & -0.342 & -0.248 & -0.158  \\ \hline
All pairwise duplications & -0.331 & -0.237 & -0.149  \\ \hline
Local composition A  &+0.658 &  +0.762 & +0.461 \\  \hline
Local composition T &+0.668 &  +0.674 & +0.551  \\ \hline
Local composition C &+0.021 & +0.039 & +0.269 \\  \hline
Local composition G &-0.149 &  +0.211 & +0.246 \\  \hline
Global composition AT &+0.052 & -0.154 & -0.219  \\
\hline \hline
\end{tabular}
\caption{\label{tab:correlation}  Correlation $\rho_C$ of the cluster fluctuations for the first  (M$_1$), the second  (M$_2$) and the third  (M$_3$) disjoint sets
of the 24 human  chromosome sequences.  The fluctuations are anticorrelated with length, gene density, inter-chromosomal and intra-chromosomal  segmental duplications, while they exhibit a positive correlation with the AT-rich regions. Very little correlation is found with the GC-rich regions and global AT composition.
Length values are shown in  the 2$^{nd}$ column of Table 1. Gene density data are taken from Refs.~\cite{Bailey,Deloukas}.  Inter- and intra-chromosomal duplications data are taken from Ref.~\cite{Bailey}. Base compositions are shown in Table 1 (respectively 4$^{th}$-7$^{th}$ columns for the whole sequence,  8$^{th}$-11$^{th}$ columns for the first 10MBases, and in Tables S1-S6 of the Supplementary Information).}
\end{table*}

\clearpage

\clearpage

\subsection*{Supplementary information}
\footnotesize
The average values  of the nucleotide contents obtained over three disjoint data sets is shown for the 24 chromosomes in Supplementary Tables S1-S6. For each set, the parameter $n$ has been varied in order to generate clusters of any lengths in random position of the sequence. The standard deviations are also reported in the same tables.  Further data, statistics and all the codes can be downloaded at the website \url{www.polito.it/noiselab} in the section \emph{Utilities}.
\footnotesize
\begin{sidewaystable}[h]
\footnotesize
\scalebox{0.8}{
\begin{tabular}{c|c|cc|cc|cc|cc||c|cc|cc|cc|cc}
 \hline \hline
  \multicolumn{9}{c}{CHROMOSOME 1}&  & \multicolumn{8}{c}{CHROMOSOME 2} \\
 \hline \hline
 & $n$  & A [\%] &  $\sigma_C$ [A] & C[\%] & $\sigma_C$ [C] &  G[\%] & $\sigma_C$ [G] &  T[\%] & $\sigma_C$ [T]  &   $n$   & A [\%] &  $\sigma_C$ [A] & C[\%] & $\sigma_C$ [C] &  G[\%] & $\sigma_C$ [G] &  T[\%] & $\sigma_C$ [T]\\
   \hline
\multirow{5}{*}{$k=1$} &2 & 27.31 & 10.11&  25.24&  11.34 & 24.81  & 9.95  &25.58  & 8.29 &2 &29.87  & 12.57  & 24.86  & 11.62  & 23.01  & 11.45  & 32.68  & 12.56\\
&4 & 26.52  &11.01 & 25.79  &10.81 & 25.58 & 10.06 & 25.15  & 9.68&4 & 28.50  & 14.19  & 24.51  & 12.90  & 22.34  & 12.43  & 29.81  & 14.30\\
&6 &24.31 &  9.62 & 26.05&  10.93 & 25.61 &  9.83  &24.72 &  9.53&6 & 26.46  & 13.42  & 24.43   &12.80   & 22.25  & 12.77  & 29.01  & 13.16\\
&8 &24.68 & 10.11 & 25.95 & 11.33 & 26.17 & 10.65  &23.59 &  9.55& 8 & 27.23  & 12.09  & 23.01  & 11.74  & 22.22 &  12.09 &  28.13  & 12.36\\
&10 & 24.48 &  8.96 & 25.89 & 10.46 & 26.18 & 10.27 & 23.77  & 8.78&10 & 26.10  & 10.89  & 23.63  & 11.12  & 22.55   &10.94 &  28.25  & 11.01\\
 \hline
\multirow{5}{*}{$k=2$}&2 & 25.94 &  4.95 & 23.64 &  4.50 & 24.24 &  5.45 & 26.18 &  5.79&2 & 27.81  &  5.29  & 21.76  &  4.57 &  23.40  &  5.81  & 27.02  &  6.36\\
&4 & 25.47 &  6.02 & 24.08  & 5.25 & 24.21 &  5.99 & 26.56  & 7.04&4 & 27.57  &  5.45  & 21.76   & 5.12  & 23.07  &  5.65  & 27.59   & 7.19\\
&6 & 25.92 &  5.66 & 23.78  & 5.12&  24.49 &  6.17 & 25.95  & 6.25&6 & 28.08  &  5.84  & 21.86   & 4.92  & 22.94  &  5.47  & 27.10  &  6.30\\
&8 & 25.88 &  5.38 & 23.80 &  4.89 & 24.51  & 5.41 & 25.81  & 5.39&8 & 28.04  &  5.38  & 21.84   & 4.70 &  23.04  &  5.80  & 27.06  &  5.84\\
&10 & 25.82 &  5.22 & 24.04 &  4.81 & 24.66  & 5.39 &  25.48 &  5.33&10 & 27.94  &  5.29  & 22.03  &  4.80  & 22.80   & 4.79  & 27.21   & 5.27\\
 \hline
\multirow{5}{*}{$k=3$}&2 & 26.62 &  6.09 & 23.00 &  5.32 & 25.04 &  6.18 & 25.84  & 5.82&2 & 30.50  &  4.45  & 20.14   & 4.42  & 20.16   & 4.12  & 29.19   & 4.72\\
&4 & 26.33 & 6.14 & 23.26  & 4.75 & 24.85 &  5.66 & 25.69  & 6.10&4 & 30.52  &  4.97  & 19.94   & 4.07  & 19.63   & 3.92  & 29.92  &  5.23\\
&6 & 26.07 &  5.41 & 23.67 &  4.94 & 24.67 &  5.27 & 25.58 &  5.33&6 & 30.29   & 4.82  & 19.84  &  3.99  & 19.68  &  3.74   &30.18  &  4.87\\
&8 & 25.84 &   5.64 & 24.44 &  5.67 & 24.64 &  5.45 & 25.34 &  5.26&8 & 30.50  &  3.82  & 19.88  &  3.81  & 19.54   & 3.21 &  30.08  &  4.28\\
&10 & 25.90 &  5.01 & 23.91 &  4.85 & 25.09 &  4.99 & 25.10 &  4.71&10 & 30.31  &  3.94  & 19.77  &  3.58  & 19.42   & 3.60  & 30.49   & 4.44\\
\hline \hline
 \multicolumn{9}{c}{CHROMOSOME 3}&  & \multicolumn{8}{c}{CHROMOSOME 4} \\
 \hline \hline
&   $n$  & A [\%] &  $\sigma_C$ [A] & C[\%] & $\sigma_C$ [C] &  G[\%] & $\sigma_C$ [G] &  T[\%] & $\sigma_C$ [T]  &   $n$   & A [\%] &  $\sigma_C$ [A] & C[\%] & $\sigma_C$ [C] &  G[\%] & $\sigma_C$ [G] &  T[\%] & $\sigma_C$ [T]\\
   \hline
\multirow{5}{*}{$k=1$}&2& 29.32 & 10.13&  22.17 &  9.85 & 22.44 &  8.30 & 28.32 &  8.32&2& 39.10  & 15.15  & 22.24  & 13.55  & 24.20  & 13.21  & 35.90  & 15.69\\
&4& 28.46 & 10.05 & 21.77 &  9.43&  21.50 &  8.29 & 28.65 &  8.52&4  &34.47   &16.56   &19.80  & 12.91   &22.86   & 13.87  & 30.28  & 17.79\\
&6& 28.72 &  9.70 & 21.61 &  8.85 & 21.93 &  8.38 & 28.08 &  8.51&6 & 31.74   &17.41   &19.41  & 12.92 &  21.92  & 13.76  & 29.21  & 18.12\\
&8& 28.59 &  9.27 & 21.35 &  8.42 & 22.15 &  8.37 & 28.08 &  8.55&8 & 31.59  & 16.35   &18.64  & 11.49  & 21.69  & 13.52  & 29.15  & 17.90\\
 &10 &28.49 &  9.01&  22.00 &  8.91 & 22.17  & 8.45 & 27.62 &  8.06&10 & 30.98  & 15.47 &  18.92 &  11.16  & 20.91  & 12.18  & 29.28  & 15.90\\
\hline
\multirow{5}{*}{$k=2$}&2& 28.68 & 6.90 & 20.84 &  4.68 & 23.39  & 7.00 & 27.46 &  8.54&2 & 30.13   & 9.37   &19.75  &  6.39  & 19.27  &  7.70   &30.85   & 9.89\\
&4& 28.43 &  5.62 & 21.10 &  4.64 & 23.13  & 5.90 & 27.73  & 6.59& 4& 31.20  &  8.37  & 18.93  &  5.45 &  18.72  &  6.98   &31.30  &  9.67\\
&6& 27.76 & 6.23 & 21.54 &  4.73&  22.86 &  5.96 & 28.03 &  6.77&6 & 31.30 &   8.21  & 18.94  &  6.12   &18.28  &  5.78  & 31.48  &  8.66\\
&8& 27.07 &  5.20&  22.35 &  4.58 & 22.64 &  5.18 & 27.95 &  6.45& 8 & 31.20 &   8.13 &  18.89  &  5.41  & 17.94   &  6.19  & 31.96  &  9.04\\
&10& 27.39 &  5.14 & 22.46 &  5.20 & 23.03  & 5.02 & 27.12  & 5.56& 10 &31.10  &  8.14  & 18.81 &   5.00   &18.12  &  5.84  & 31.97  &  9.13\\
\hline
\multirow{5}{*}{$k=3$}&2& 30.73  & 6.31 & 19.21 &  4.50 & 19.99 &  4.27 & 30.37 &  5.37&2& 31.09  &  5.36   &19.86   & 5.24  & 20.44   & 6.64  & 29.08  &  6.77\\
&4& 31.25 &  5.96 & 18.87 &  4.17 & 19.48 &  3.64 & 30.65 &  4.90& 4& 30.77   & 5.61  & 19.27  &  4.96  & 20.04   & 5.88  & 29.93   & 6.42\\
&6& 30.67 &  4.88 & 19.29 &  4.02 & 19.57  & 4.32 & 30.47 &  5.21& 6& 31.17   & 5.90 &  19.24  &  4.72  & 20.10  &  5.11   &29.49   & 6.37\\
&8& 30.84 &  4.35 & 19.04 &  3.27 & 19.56 &  3.98 & 30.56 &  5.85& 8& 31.42  &  5.57 &  19.08   & 4.11  & 20.04   & 5.22   &29.46   & 5.89\\
&10& 30.12 &  5.09 & 19.46  & 3.70 & 19.94 &  4.20 & 30.49 &  5.11& 10& 31.25 &   5.16  & 19.21   & 4.25  & 20.01   & 4.82  & 29.53   & 5.89\\
 \hline \hline
\end{tabular}}
{\footnotesize TABLE S1. Average Nucleotide Composition of the Clusters.  Base composition and standard deviation of nucleotide \% of A  (3$^{rd}$ and 4$^{th}$ columns), \%  of  C  (5$^{th}$ and 6$^{th}$ columns), \%  of G (7$^{th}$ and 8$^{th}$ columns), \%  of  T (9$^{th}$ and 10$^{th}$ columns).   Different $k's$ refer to the first  (M$_1$), second  (M$_2$) and  third  (M$_3$) disjoint sets
of the chromosomes 1, 2, 3, 4.  Different values of $n$  are shown also.
}
\end{sidewaystable}

\clearpage
\begin{sidewaystable}[h]
\footnotesize
\scalebox{0.8}{
\begin{tabular}{c|c|cc|cc|cc|cc||c|cc|cc|cc|cc}
\hline \hline
  \multicolumn{9}{c}{CHROMOSOME 5}&  & \multicolumn{8}{c}{CHROMOSOME 6} \\
 \hline \hline
  &   $n$   & A [\%] &  $\sigma_C$ [A] & C[\%] & $\sigma_C$ [C] &  G[\%] & $\sigma_C$ [G] &  T[\%] & $\sigma_C$ [T]  &  $n$   & A [\%] &  $\sigma_C$ [A] & C[\%] & $\sigma_C$ [C] &  G[\%] & $\sigma_C$ [G] &  T[\%] & $\sigma_C$ [T]\\
   \hline
\multirow{5}{*}{$k=1$}  &2 & 38.67  & 16.10  & 26.08  & 13.44  & 20.28  & 11.83  & 42.62  & 13.40&2 & 30.20  &  9.93  & 21.79  &  7.09  & 21.10   & 6.63  & 28.49   & 8.97\\
 & 4 &29.97  & 19.75  & 24.86  & 14.32  & 19.70  & 12.76  & 36.43   & 16.69&4 & 29.97  &  9.07  & 21.23  &  6.33  & 21.73   & 7.42   &28.27   & 8.71\\
 &6 & 27.90 &  19.23  & 24.21   &13.69  & 19.11  & 11.78  & 33.24  & 17.76&6 & 29.00  &  8.84  & 22.04  &  7.86   &21.67   & 7.58   &27.75   & 8.37\\
 &8 & 26.77  & 16.82  &  24.07  & 12.89  & 18.83  & 11.45  & 32.46   &16.22&8 & 28.69  &  8.83  & 22.15  &  8.10  & 21.44   & 6.50  & 28.30   & 8.17\\
 & 10 & 26.72  & 16.00  &  23.66  & 12.64  & 19.19  & 11.29  & 31.19   &15.47&10 & 28.35  &  7.90 &  22.54  &  7.63  & 21.03  &  7.12  & 28.33   & 7.60\\
   \hline
\multirow{5}{*}{$k=2$} & 2& 34.90  & 16.10  & 21.89   &13.00   &23.40  & 10.37  & 33.17  & 15.08&2 & 28.77   & 5.30  & 21.47   & 4.72  & 21.11  &  3.97   &28.65   & 5.46\\
 &4 & 32.36  & 16.35  & 21.60  &  12.87  & 23.98  & 12.58  & 30.32  & 16.83&4 & 28.59  &  5.14  & 21.33   & 4.26  & 21.46   & 3.85  & 28.62   &  5.13\\
 & 6 &28.81  & 15.12  & 22.37  & 12.87   &21.98   &11.66  & 29.16   &16.07&6 & 28.45   & 5.77 &  21.66   & 5.25  & 21.47   & 4.44  & 28.38   & 5.03\\
 &  8&29.67  & 14.59  & 21.70  &  12.20 &  22.39  & 11.59  & 27.21 &  14.74&8 & 28.48   & 5.60   &21.36  &  5.65  & 21.43   & 3.79  & 28.67   & 5.06\\
 & 10 &29.18  & 12.77  & 21.27  &  10.16 &  21.99   & 9.79  & 28.06  & 13.01&10 & 28.64  &  5.11   &21.20  &  3.88  & 21.21   & 3.64  & 28.85   & 4.77\\
   \hline
\multirow{5}{*}{$k=3$} &2  &30.18  & 10.32  & 19.67   & 8.34  & 23.38   & 9.83   &28.88  & 12.60& 2 &30.73   & 5.02  & 19.02   & 4.47  & 19.94  &  4.74   &30.31  &  6.04\\
 &4 & 30.94  & 11.06  & 18.89   & 7.70  & 23.18  & 10.12  & 28.08   &12.35& 4 &30.67  &  4.97  & 19.47   & 4.60  & 19.99   & 4.22  & 29.87  &  6.10\\
 &6 & 30.71  & 10.08  & 19.16   & 7.18  & 22.21   & 9.14  & 28.06  & 11.45&6 & 30.33  &  4.60  & 19.85 &   4.31  & 20.26  &  3.60   &29.55   & 5.22\\
 &8 & 30.69  &  9.52  & 18.75   & 6.80  &  21.84   & 9.15   & 28.89  & 11.13&8 & 30.39  &  4.59  &19.59    &4.03 &  20.05   & 3.67   &29.97   & 4.97\\
 &10 & 30.44  &  9.13  & 18.89  &  6.49  & 21.70  &  8.72  & 29.13  & 10.66&10 & 30.50  &  4.64   &19.64   & 3.84  & 20.13  &  4.30  & 29.73  &  5.36\\
 \hline \hline
  \multicolumn{9}{c}{CHROMOSOME 7}&  & \multicolumn{8}{c}{CHROMOSOME 8} \\
 \hline \hline
  &   $n$  & A [\%] &  $\sigma_C$ [A] & C[\%] & $\sigma_C$ [C] &  G[\%] & $\sigma_C$ [G] &  T[\%] & $\sigma_C$ [T]  &  $n$   & A [\%] &  $\sigma_C$ [A] & C[\%] & $\sigma_C$ [C] &  G[\%] & $\sigma_C$ [G] &  T[\%] & $\sigma_C$ [T]\\
   \hline
\multirow{5}{*}{$k=1$} & 2 & 29.06   & 8.12  & 21.65   & 8.09  & 26.51  & 10.55  & 23.22   & 9.51&2 &29.46  & 4.78 & 20.44   &4.08 & 21.30 &  4.23 & 28.79  & 4.65\\
 &4 & 28.85   & 8.58  & 21.79  &  8.32  & 27.09  & 11.14  & 22.93  & 10.21&4 & 29.51  & 4.89 & 20.61  & 3.88 & 20.85 &  4.33 & 29.03 &  4.91\\
 &6 & 27.51   & 8.34  & 22.31   & 8.85  & 26.65  & 10.31  & 23.65  &  9.47&6 & 29.18 &  4.85 & 20.85 &  3.88 & 21.23  & 4.10  &28.74 &  4.45\\
 &8 & 26.88   & 8.22  & 23.33  & 10.23  & 25.85  & 10.49  & 24.07  &  8.90&8 & 28.74 &  5.21 & 21.17 &  4.22 & 21.94  & 4.97 & 28.15 &  4.72\\
 &10 & 26.38   & 7.91  & 23.97  &  9.78  & 25.28   & 9.83 &  24.59   & 8.17&10 & 28.46  & 4.67 & 20.91&   3.88 & 22.21  & 4.88 & 28.43  & 4.45\\
   \hline
\multirow{5}{*}{$k=2$} &2 & 32.16  &  8.29  & 19.75  &  6.25  & 19.37  &  7.27  & 29.04  &  8.29&2& 30.33  & 5.23 & 20.37 &  5.01 & 21.02 &  4.78 & 28.28 &  5.36\\
 &4  &31.99  &  6.91 &  18.97  &  4.99  & 19.38   & 6.41  & 29.94   & 7.62&4& 30.44  & 4.66 & 20.30 &  4.35 & 20.65 &  5.24 & 28.61 &  4.99\\
 &6 & 31.87   & 6.03   &18.93  &  4.83  & 18.84  &  5.51  & 30.49  &  6.76&6 & 30.28  & 4.74 & 20.53 &  4.66 & 20.44  & 4.32 & 28.75 &  4.90\\
 &8 & 31.43   & 5.29  & 18.76   & 4.20  & 18.65   & 5.27  & 31.28  &  6.69&8& 30.43 &  4.68 & 20.29 &  4.38 & 20.40 &  4.41 &  28.88  & 5.23\\
 &10 & 31.40   & 5.45  & 18.70   & 4.13  & 18.59   & 4.47   &31.31  &  6.11&10& 30.15 &  4.88 & 20.36 &  4.63 & 20.52 &  4.80 & 28.97  & 4.51\\
   \hline
\multirow{5}{*}{$k=3$} &2  &30.60   & 6.82  & 20.99   & 4.29  &  19.99   & 4.94  & 28.43  &  6.46&2& 27.97 &   6.08 & 22.15 &  4.99&  21.72 &  4.42 & 28.16 &  6.32\\
 &4  &29.28  &  7.22  & 21.43   & 4.72  & 20.28   & 5.10   &29.01   & 7.62&4& 27.87 &  5.33 & 22.38 &  5.20&  21.63 &  4.41 & 28.12 &  5.59\\
 &6  &29.19   & 5.73  & 21.07   & 4.05  & 20.73   & 4.90  & 29.01   & 6.28&6& 28.00 &  4.78 & 21.75 &  4.48 & 22.08  & 4.14 & 28.18 &  5.45\\
 &8 & 29.57  &  5.63  & 20.86   & 3.96  &  20.67   & 4.94  & 28.91  &  6.07&8& 28.27 &  4.71 & 21.95 &  4.26 &  21.73 &  4.21&  28.06 &  5.15\\
 &10 & 29.48  &  5.45  & 20.67   & 3.97  & 20.70   & 4.45  & 29.15   & 5.68&10& 27.69 &  4.63 & 22.01 &  4.31 & 22.13 &  4.04 & 28.17 &  4.90\\
\hline \hline
\end{tabular}}
{\footnotesize  TABLE S2. Same as Table S1 but for the chromosomes 5, 6, 7, 8.}
\end{sidewaystable}

\clearpage

\begin{sidewaystable}[h]
\footnotesize
\scalebox{0.8}{
\begin{tabular}{c|c|cc|cc|cc|cc||c|cc|cc|cc|cc}
\hline \hline
  \multicolumn{9}{c}{CHROMOSOME 9}&  & \multicolumn{8}{c}{CHROMOSOME 10} \\
 \hline \hline
   &  $n$   & A [\%] &  $\sigma_C$ [A] & C[\%] & $\sigma_C$ [C] &  G[\%] & $\sigma_C$ [G] &  T[\%] & $\sigma_C$ [T]  &  $n$   & A [\%] &  $\sigma_C$ [A] & C[\%] & $\sigma_C$ [C] &  G[\%] & $\sigma_C$ [G] &  T[\%] & $\sigma_C$ [T]\\
   \hline
\multirow{5}{*}{$k=1$} &2& 28.47 &  5.05 & 21.72 &  4.61 & 20.93 &  4.92 & 28.87 &  5.52&2&31.61 &  9.77&  19.73 &  8.05  &19.88&   7.08  &29.08 &  8.92\\
 &4& 27.91 &  6.49 & 21.83 &  4.97 & 21.38 &  4.36 & 28.76  & 5.23&4& 31.15  &10.84  &19.94  & 9.04 & 19.47 &  7.52 & 29.44&   9.38\\
 &6& 27.84 &  6.07 & 22.07 &  6.37 & 21.94 &  5.03 & 28.21 &  5.71 &6 &31.59  & 9.73  &20.05  & 7.63  &19.56&   8.08 & 29.46 & 10.13\\
 &8& 27.99 &  6.19 & 22.19 &  7.00&  21.46 &  5.65 & 28.29  & 5.17 &8& 30.85  &10.08&  20.34&   8.58  &18.82&   7.72 & 30.12  & 9.73\\
 &10& 27.96&   5.03 & 22.05 &  6.37 & 21.89 &  5.59 & 28.04  & 4.92 &10& 31.08 &  9.55 & 20.40  & 8.43 & 18.10  & 7.38 & 30.53 &  8.92\\
   \hline
\multirow{5}{*}{$k=2$} &2& 29.60  & 5.12 & 20.23 &  4.27 & 20.20  & 4.67 & 29.98  & 6.23&2&29.87  &11.56  &19.91  & 7.52&  24.02 &  8.49&  27.02&  11.13\\
 &4& 29.36  & 6.14 & 20.75 &  4.63 & 19.51 &  4.61 & 30.38  & 6.14 &4&30.28  &11.01 & 19.59  & 7.60&  22.90 &  8.37 & 27.98&  10.54\\
 &6& 29.55 &  4.68 & 20.19 &  3.94 & 19.80  & 4.24 & 30.45  & 5.58&6&30.38&  10.44  &19.42  & 7.48 & 21.99 &  8.48 & 29.04 & 10.80\\
 &8& 29.68  & 4.73 & 20.13 &  3.81 & 19.73 &  3.69 & 30.46 &  4.79 &8& 30.99 &  9.40 & 19.22 &  6.53 & 21.17 &  7.71 & 28.98 &  9.39\\
 &10& 29.47 & 4.48 & 20.34 &  3.77 & 19.74  & 3.16  &30.45 &  4.72&10&30.63 &  8.43  &19.68 &  5.99 & 21.00 &  7.43&  28.87 &  8.91\\
   \hline
\multirow{5}{*}{$k=3$} &2 &30.03  & 5.76 & 20.57 &  4.51 & 21.83 &  5.19 & 27.78 &  4.77 &2 &27.22 &  9.61 & 21.98  & 7.65  &24.82 &  9.16  &26.85  &11.75\\
 &4 &29.85  & 5.06&  20.59 &  3.97 & 21.60 &  4.51 & 27.96  & 4.98 & 4&27.53 &  9.86&  21.86&   7.66 & 24.90 &  9.57&  25.90  &11.44\\
 &6& 29.56  & 4.86 & 20.83 &  3.98 & 21.58 &  4.42 & 28.03 &  4.81&6&28.31 &  9.00 & 21.34 &  7.05 & 24.67  & 8.44 & 25.85 &  9.76\\
 &8& 29.31  & 4.58 & 20.97 &  3.82 & 21.50 &  3.80 & 28.23 &  4.47 &8&28.18 &  8.98&  22.08 &  7.09 & 23.66  & 7.41 & 26.24  & 9.08\\
 &10& 29.14 &  4.69&  21.10 &  3.72 & 21.55 &  3.58 & 28.22 &  4.00& 10&27.46 &  8.53 & 22.20&   6.99&  23.74 &  8.08&  26.59 &  9.50\\
 \hline \hline
 \multicolumn{9}{c}{CHROMOSOME 11}&  & \multicolumn{8}{c}{CHROMOSOME 12}\\
\hline \hline
   &  $n$   & A [\%] &  $\sigma_C$ [A] & C[\%] & $\sigma_C$ [C] &  G[\%] & $\sigma_C$ [G] &  T[\%] & $\sigma_C$ [T]  &  $n$  & A [\%] &  $\sigma_C$ [A] & C[\%] & $\sigma_C$ [C] &  G[\%] & $\sigma_C$ [G] &  T[\%] & $\sigma_C$ [T]\\
 \hline
\multirow{5}{*}{$k=1$} &2&  29.26 &   9.87 &  21.99 &   7.82 &  24.40 &   11.32&   27.47  & 10.11&2& 34.76 &  14.95  & 24.49 &  13.32 &  25.09  & 13.05  & 34.26 &  13.25\\
 &4&  28.97  &  9.12 &  22.23 &   8.83 &  25.08 &  11.69  & 26.85 &  10.87&4& 30.66 &  17.09 &  22.85&   14.86 &  24.19&   14.13&   29.62&   15.63\\
 &6&  26.63 &   8.87&   23.26 &   9.46 &  24.78 &  12.88&    26.66 &  11.31&6& 30.02  & 16.87  & 21.91 &  14.19  & 24.09&   14.12  & 27.59 &  15.58\\
 &8&  26.28 &   9.23 &  23.93 &  10.76 &  23.93  & 11.97 &  26.28 &  10.33&8& 27.79 &  14.28&   23.12 &  12.79  & 22.88 &  13.10  & 27.62 &  14.11\\
 &10&  25.65 &   9.03  & 23.52 &  10.98  & 24.83 &  13.14&   26.23 &  10.48&10& 26.73 &  13.38&   23.06 &  12.73  & 23.35 &  13.05 &  27.37 &  13.31\\
 \hline
\multirow{5}{*}{$k=2$} &2&  27.26 &  11.01  & 21.67 &   7.27&  21.50  & 8.77 & 30.41  &13.20&2& 28.06  & 10.73  & 20.87 &   7.88  & 21.15   & 9.12  & 30.17  & 11.40\\
 &4&  27.02 &  10.15  & 21.86&    8.03&  20.83 &  8.05&  30.48&  11.75&4& 28.45 &  10.97 &  20.27  &  7.68 &  20.31 &   9.67&   30.97 &  11.56\\
 &6&  27.48 &  10.23  & 21.61 &   7.74&  20.90&   8.36&  30.18 & 11.51&6& 28.98&   10.18 &  20.31&    7.16&   19.92  &  7.94 &  31.13&   10.87\\
 &8 &  27.58 &   9.68 &  21.78  &  7.39 & 21.12 &  8.18&  29.52&  11.08&8& 29.41   & 9.46 &  20.19  &  6.58  & 19.80 &   7.58 &  30.61  & 10.08\\
 &10& 27.60 &   9.69 &  21.96 &   7.93&  21.01&   7.43&  29.71  &10.35&10& 29.06  &  9.23 &  20.36  &  6.67&   19.83  &  7.71&   30.75 &   9.65\\
 \hline
\multirow{5}{*}{$k=3$} &2&  30.18 &  11.93  & 20.84 &  11.12&   21.56&   10.35  & 29.10  & 11.36&2& 29.75 &  11.94  & 20.78  &  8.79 &  21.48 &   9.97  & 29.38  & 13.11\\
 &4&  30.14  & 11.36 &  20.66&   10.53 &  20.14  &  9.56  & 29.49&   11.32&4& 29.85&   10.84&   20.83  &  8.62 &  20.24&    8.97 &  30.15&   12.35\\
 &6&  30.55 &  10.57 &  20.02 &   9.98 &  20.14  &  9.46&   29.60 &  11.05&6& 29.53 &  10.20 &  20.62&    8.75  & 19.72 &   8.66 &  30.59 &  11.18\\
 &8&  29.78 &  10.89 &  20.87&   10.23  & 19.24   & 9.04  & 30.23 &  10.77&8& 30.83&    9.56 &  20.07  &  7.30 &  19.40 &   8.59 &  29.88 &  10.96\\
 &10&  30.02 &   9.48 &  20.36 &   8.70 &  19.30&    8.14  & 30.32  &  9.51&10& 30.30 &   9.38 &  20.05 &  7.03 &  19.33 &   8.09&   30.33&   10.43\\
 \hline \hline
\end{tabular}}
{\footnotesize  TABLE S3. Same as Table 1 but for the chromosomes 9, 10, 11, 12.}
\end{sidewaystable}

\clearpage

\begin{sidewaystable}[h]
\footnotesize
\scalebox{0.8}{
\begin{tabular}{c|c|cc|cc|cc|cc||c|cc|cc|cc|cc}
\hline \hline
  \multicolumn{9}{c}{CHROMOSOME 13}&  & \multicolumn{8}{c}{CHROMOSOME 14}\\
 \hline \hline
   &  $n$   & A [\%] &  $\sigma_C$ [A] & C[\%] & $\sigma_C$ [C] &  G[\%] & $\sigma_C$ [G] &  T[\%] & $\sigma_C$ [T]  &   $n$   & A [\%] &  $\sigma_C$ [A] & C[\%] & $\sigma_C$ [C] &  G[\%] & $\sigma_C$ [G] &  T[\%] & $\sigma_C$ [T]\\
   \hline
\multirow{5}{*}{$k=1$}  &2&     40.65  &  14.23&    22.09&    12.40 &   23.04 &   13.69&    41.30 &   14.70  &2&  41.94  &  15.83  &  20.20 &   11.03  &  23.04  &  13.01 &   40.63 &   17.09\\
&4&       33.94&    17.12  &  20.95  &  12.01 &   21.82 &   13.78 &   33.88 &   18.11    &4&  35.84  &  19.53  &  20.02 &   13.06 &   21.92  &  13.43 &   30.95 &   19.26\\
 &6&       30.35 &   16.75&    20.83  &  12.45&    21.90 &   12.49 &   30.88  &  17.42  &6&  34.11  &  17.06  &  18.23 &   11.65 &   23.30  &  12.87 &   27.56  &  18.13\\
&8&       29.31 &   15.73 &   21.22 &   11.54 &   21.18&    12.66 &   30.10 &   16.99    &8& 32.72  &  17.23 &   18.49  &  11.55&    23.28 &   12.98&    27.16&    17.71\\
&10&   29.54  &  14.63 &   21.12 &   11.21 &   21.56 &   11.51&    29.00&    15.72     &10&  31.44 &   15.34&    19.18  &  11.52 &   22.63  &  12.56  &  27.63 &   15.97\\
 \hline
\multirow{5}{*}{$k=2$}  &2&  42.73  &  17.08 &   21.84 &   11.77&    22.74  &  13.27&    43.58  &  15.19  &2&  40.24 &   16.10&    23.25 &   14.31&    21.61&    13.92 &   40.57 &   12.29\\
 &4&  35.36  &  19.85  &  20.15 &   13.49&    20.58 &   12.48 &   35.63  &  17.46 &4&  33.37 &   18.24 &   21.03 &   13.95 &   20.33  &  14.74 &   34.19&    18.21\\
 &6&  35.00&    20.06&    18.97 &   11.70 &   21.21 &   12.20&    30.10  &  19.56&6& 30.48  &  19.31 &   21.66  &  15.06  &  19.76  &  14.53 &   31.55 &   18.48 \\
 &8 & 32.55 &   17.25 &   18.68  &  10.58  &  21.90&    11.83&    28.91 &   16.84 &8& 29.30 &   16.72 &   21.21 &   12.93 &   19.25 &   12.89&    31.43  &  16.17\\
 &10&  31.63&    17.27 &   19.06&    10.87 &   21.78 &   10.49  &  28.56 &   16.23&10& 29.92  &  15.80  &  20.79 &   11.81&    19.14  &  11.12 &   30.63 &   14.70 \\
 \hline
\multirow{5}{*}{$k=3$}  &2& 41.87  &  14.57&    20.52 &   12.37 &   25.57&    14.60 &   37.41 &   13.91 &2&  41.55  &  15.36  &  22.11 &   14.66 &   23.34  &  12.15&    37.42 &   12.47\\
 &4& 34.21&    17.35 &   19.82 &   14.46&    24.50  &  15.43 &   28.82 &   18.55  &4& 36.90 &   18.39 &   20.44  &  15.14&    22.74  &  13.00 &   32.29  &  17.48\\
 &6& 32.78  &  16.31&    19.16&    13.24 &   24.29 &   13.45 &   27.20 &   16.99 &6& 33.79 &   17.42  &  19.15 &   13.34 &   22.07&    12.46&    30.30 &   15.69 \\
 &8& 32.22 &   15.61&    19.01 &   12.64 &   24.26  &  13.08 &   26.31 &   16.03 &8& 33.86 &   16.52&    18.09&    12.17 &   20.97 &   11.97  &  28.82  &  15.26 \\
 &10&  30.65  &  14.53&    19.87 &   11.66&    22.75  &  12.32 &   27.88  &  15.07 &10&  33.24&    15.21 &   18.15&    10.94&    20.18 &   10.81&    29.37&    14.49\\
 \hline \hline
  \multicolumn{9}{c}{CHROMOSOME 15}&  & \multicolumn{8}{c}{CHROMOSOME 16}\\
 \hline \hline
   &  $n$  & A [\%] &  $\sigma_C$ [A] & C[\%] & $\sigma_C$ [C] &  G[\%] & $\sigma_C$ [G] &  T[\%] & $\sigma_C$ [T]  &   $n$   & A [\%] &  $\sigma_C$ [A] & C[\%] & $\sigma_C$ [C] &  G[\%] & $\sigma_C$ [G] &  T[\%] & $\sigma_C$ [T]\\
   \hline
\multirow{5}{*}{$k=1$}  &2&     40.84 & 14.37&  22.25  &13.19  &23.12  &13.67 & 40.13 & 14.71          &2&  34.88  &13.87 & 26.22  &14.91&  23.30  &13.63  &37.54 & 14.59 \\
 &4&     32.64&  16.55 & 21.69&  14.08 & 20.74  &12.54 & 33.00  &16.61        &4&  29.17&  16.31 & 24.30 & 15.60&  22.89 & 15.77 & 31.49 & 15.69\\
 &6&     28.66  &17.57&  21.93 & 13.45 & 20.68&  12.84 & 30.43 & 16.80         &6& 26.91 & 17.84&  24.44  &15.90  &21.56&  15.33 & 29.44 & 16.86 \\
 &8&        27.51&  16.52&  22.48  &13.72 & 20.39 & 12.91 & 30.04&  16.31         &8& 26.43 & 16.97 & 23.79 & 15.97&  22.74  &15.51 & 27.78 & 16.10\\
 &10&   28.58 & 15.42 & 22.23&  12.96  &20.69 & 11.69&  29.10  &14.67            &10&  26.67&  15.58&  23.77 & 15.40 & 23.00  &14.99&  26.56 & 14.94\\
 \hline
\multirow{5}{*}{$k=2$}  &2&  40.26 & 14.52&  22.73&  16.03&  23.13  &14.63 & 36.79  &11.11 &2&  38.33 & 14.90  &24.35  &13.47 & 23.99&  12.90&  38.60&  15.42\\
 &4& 32.34  &18.19 & 22.98&  16.33  &21.29 & 14.48  &31.79 & 16.12  &4& 31.94 & 17.61 & 22.15 & 13.18&  22.26  &13.36&  33.44 & 18.60 \\
 &6& 31.23&  17.22 & 21.30 & 14.22 & 22.31 & 13.14  &28.10 & 16.50&6&  29.28  &18.50 & 22.01 & 14.21&  22.22 & 14.79&  30.41&  19.57 \\
 &8 & 30.73&  15.97 & 21.10  &13.49&  21.56&  12.23  &27.67 & 14.99&8& 27.47&  17.88&  22.30 & 13.95  &21.70 & 15.03&  29.83&  18.42\\
 &10&  30.12&  14.71&  20.70&  11.97 & 21.94&  11.71&  27.72 & 14.31&10&  26.21 & 17.27&  22.85&  14.23 & 21.21 & 14.77 & 30.13 & 17.77\\
 \hline
\multirow{5}{*}{$k=3$}  &2&  37.32 & 10.29  &23.57 & 14.73&  21.14 & 13.91  &42.04  &15.39&2& 33.14  &12.52  &22.66 & 10.96 & 21.99  &12.25  &34.38 & 15.95 \\
 &4&  30.90 & 18.55 & 23.68&  15.16 & 19.56&  13.97 & 35.87  &18.51 &4& 28.95  &13.80  &22.68 & 11.12  &21.02  &11.42 & 31.08  &15.86\\
 &6& 26.78  &15.31 & 22.30  &12.87 & 19.83&  14.77  &34.00&  17.49 &6& 26.82&  13.90&  23.25  &12.02&  20.76&  11.06  &30.52  &14.57 \\
 &8&  27.56&  14.08 & 21.79 & 11.45  &19.29 & 11.82&  32.83 & 15.67&8&  26.30 & 13.54 & 23.29  &11.90&  20.48  &11.38 & 30.87 & 14.32\\
 &10& 27.57 & 13.71  &21.65  &11.93&  20.55  &13.03 & 31.30&  15.35  &10&  27.63&  12.70 & 21.72 & 10.19 & 21.03&  10.32 & 30.07  &13.09\\
 \hline \hline
\end{tabular}}
{\footnotesize  TABLE S4. Same as Table 1 but for the chromosomes 13, 14, 15, 16.}
\end{sidewaystable}

\clearpage

\begin{sidewaystable}[h]
\footnotesize
\scalebox{0.8}{
\begin{tabular}{c|c|cc|cc|cc|cc||c|cc|cc|cc|cc}
\hline \hline
  \multicolumn{9}{c}{CHROMOSOME 17}&  & \multicolumn{8}{c}{CHROMOSOME 18}\\
 \hline \hline
   &  $n$  & A [\%] &  $\sigma_C$ [A] & C[\%] & $\sigma_C$ [C] &  G[\%] & $\sigma_C$ [G] &  T[\%] & $\sigma_C$ [T]  &  $n$  & A [\%] &  $\sigma_C$ [A] & C[\%] & $\sigma_C$ [C] &  G[\%] & $\sigma_C$ [G] &  T[\%] & $\sigma_C$ [T]\\
   \hline
\multirow{5}{*}{$k=1$} &2&    25.71 &  5.54  &24.38&   5.28 & 24.65 &  4.95 & 24.90  & 5.20 &2&  42.55 & 14.55&  22.81 & 13.48  &25.01 & 14.13 & 40.47 & 13.18\\
 &4&     25.36  & 5.06&  24.80  & 5.17  &24.73  & 4.95 & 24.87  & 5.67   &4&  36.73&  20.02 & 21.26 & 13.98 & 21.86 & 14.10 & 31.97&  18.47\\
 &6&      26.11 &  5.08 & 24.49 &  4.16 & 24.23   &4.08  &24.95 &  4.30  &6& 33.41 & 18.97 & 20.95 & 13.48  &21.45&  13.36  &28.45 & 17.62 \\
 &8&    25.49  & 4.55 & 24.52  & 4.19&  24.35  & 4.18 & 25.43 &  4.50    &8& 32.92&  16.95  &20.01 & 10.97 & 21.56 & 11.11 & 27.60 & 16.08\\
 &10&     25.35&   4.09  &24.91 &  4.04 & 24.65 &  4.14&  24.90 &  3.96    &10& 32.53&  15.32  &19.32  &10.33 & 21.74  &10.80 & 27.25 & 15.02 \\
 \hline
\multirow{5}{*}{$k=2$} &2&  27.62  & 5.13&  21.78 &  5.09 & 22.77&   5.48 & 27.83  & 5.36 &2&  31.93 & 11.44 & 22.82  &10.61&  21.59 & 10.29&  31.88  &12.57\\
 &4& 27.63&   4.99 & 22.07  & 5.00 & 22.28  & 4.50 & 28.02  & 4.99  &4& 29.46 & 11.69 & 21.76  & 9.87 & 21.95 &  9.81 & 29.95  &12.44 \\
 &6&  27.95   &5.91 & 22.24  & 5.39 & 21.96 &  5.11&  27.84 &  4.96&6& 28.09 & 11.62 & 21.76  & 9.12 & 21.62&   8.95  &29.66  &12.16 \\
 &8 &  27.62 &  4.73 & 22.16 &  4.45 & 22.20 &  4.36 & 28.02&   4.94&8& 28.07 & 11.33&  21.32  & 9.83&  21.95&  10.62 & 29.45 & 12.58\\
 &10& 27.57  & 4.64 & 22.17 &  4.19&  22.27&   4.01 & 27.99 &  4.34 &10& 27.79  &11.30&  21.26 &  9.22 & 21.42   &9.55 & 29.65 & 12.03 \\
 \hline
\multirow{5}{*}{$k=3$} &2&28.52 &  6.19&  22.41&   4.87 & 21.39 &  4.51&  27.67  & 5.12 &2& 31.20 &  7.92 & 19.97 &  6.75  &19.63   &5.92 & 29.20 &  7.57 \\
 &4& 28.19  & 5.44&  22.51 &  4.84&  21.68 &  4.80&  27.62 &  4.97  &4& 31.45 &  8.15 & 20.03 &  5.94  &19.84 &  5.90&  29.06 &  7.43\\
 &6& 27.83  & 5.20 & 22.68 &  4.89 & 21.71  & 4.45 & 27.78 &  4.89 &6& 30.93&   6.62  &19.89 &  5.88 & 20.03  & 5.79 & 29.30 &  7.17 \\
 &8&  27.86 &  5.13  &22.46 &  4.63&  21.74  & 4.43 & 27.94 &  4.70&8& 31.34 &  6.30 & 19.54 &  5.79 & 19.90  & 5.49  &29.36 &  6.48 \\
 &10&  27.61  & 4.83 & 22.71  & 4.22 & 21.89 &  4.03 & 27.80 &  4.42 &10& 31.25 &  5.79 & 19.48 &  5.72 & 19.64  & 5.51  &29.63 &  6.21 \\
 \hline \hline
  \multicolumn{9}{c}{CHROMOSOME 19}&  & \multicolumn{8}{c}{CHROMOSOME 20}\\
 \hline \hline
   & $n$   & A [\%] &  $\sigma_C$ [A] & C[\%] & $\sigma_C$ [C] &  G[\%] & $\sigma_C$ [G] &  T[\%] & $\sigma_C$ [T]  &   $n$   & A [\%] &  $\sigma_C$ [A] & C[\%] & $\sigma_C$ [C] &  G[\%] & $\sigma_C$ [G] &  T[\%] & $\sigma_C$ [T]\\
   \hline
 \multirow{5}{*}{$k=1$}&2& 38.68 &  16.90 & 25.41 & 13.50 & 23.06 &  11.51 & 38.36 & 14.56&2 &35.49 & 13.10 & 24.93 & 12.17 & 21.37 & 12.11 & 41.67 & 15.11\\
 &4& 32.65 & 17.90 & 21.61 & 14.88 & 23.15 & 13.99 & 30.64 & 17.10&4 & 29.01 & 17.31 & 24.09 & 13.86 & 19.02 &  14.33 & 36.45&  18.70\\
 &6& 30.43&  18.05 & 20.61 & 15.50 & 24.70 & 15.61 & 26.28 & 17.62&6 & 27.49 & 16.91&  23.23 & 13.43 & 19.70 & 14.02 & 32.51 & 18.17\\
 &8& 29.95 & 18.15 & 20.73 & 15.20 & 24.61 & 15.06 & 25.99 & 17.05&8 & 25.65 &  16.25 & 24.03 & 13.52 &  19.20 & 14.16 & 32.86 & 17.67\\
 &10& 29.30&  16.57&  21.25 & 14.76 & 24.64 & 15.24 & 25.51&  16.54&10 & 26.47 & 15.40&  22.98 & 12.55 & 20.00 & 13.28 & 31.47 & 16.89\\
   \hline
\multirow{5}{*}{$k=2$} &2& 38.61  &14.65 & 20.05 & 11.18 & 25.03 & 13.78 & 36.39 & 14.68&2& 37.88 & 14.25 & 23.84 & 13.07 & 22.79 & 11.18 & 38.59 & 14.92\\
 &4 &34.50 & 16.28 & 19.97 & 12.87 & 25.59&  14.52 & 28.19 & 17.72&4& 31.75&  18.28 & 22.38 & 14.30 & 20.80 & 12.67 & 33.24 & 18.42\\
 &6& 30.68 & 16.89 & 21.23 & 13.41 & 25.19 & 15.09 & 26.28 & 17.61&6& 29.21 & 17.20 & 21.65 & 14.26 & 21.59 & 14.24 & 31.39 & 18.24\\
 &8 & 29.23 & 15.55&  21.85 & 12.49 & 24.95 & 13.11 & 25.28 & 15.67&8& 27.95 & 13.06 & 22.28 & 11.21 & 23.55 & 11.85&  26.91 & 13.73\\
 &10& 27.52 & 15.23 & 23.16 & 12.88 & 25.04 & 12.76 & 24.72 & 14.68&10& 28.68 & 15.74&  21.82 & 13.93 & 20.67 & 13.14 & 29.55 & 15.70\\
   \hline
\multirow{5}{*}{$k=3$} &2& 35.86 & 13.33 & 25.37 & 13.36 & 23.27 & 13.08 & 38.80 & 15.55&2& 33.43 & 12.67 & 22.50&  10.70 & 23.62 &  9.91 & 32.60 & 15.17\\
 &4& 29.55 & 16.31 & 24.42 & 14.64 & 22.30 & 14.52 & 31.97 & 17.94&4& 31.75&  18.28 & 22.38 & 14.30 & 20.80 & 12.67 & 33.24 & 18.42\\
 &6& 26.12 & 17.33 & 25.32 & 15.54 & 22.15  &15.52 & 29.65 & 17.41&6& 28.45 & 14.39 & 21.95 & 11.96 & 22.89 & 12.12&  28.57 &  14.46\\
 &8& 25.09 & 16.78 & 25.20 & 15.10 & 22.89 & 14.72 & 28.67 & 17.03&8& 27.95 & 13.06 & 22.28 & 11.21 & 23.55 & 11.85&  26.91 & 13.73\\
 &10& 24.33 & 14.52 & 25.66 & 14.61 & 24.01 & 14.61 & 26.30 & 14.77&10& 27.78 & 12.94  &22.06 & 11.55 & 23.88 & 11.58 & 26.63 & 13.24\\
 \hline \hline
\end{tabular}}
{\footnotesize  TABLE S5. Same as Table 1 but for the chromosomes 17, 18, 19, 20.}
\end{sidewaystable}

\clearpage

\begin{sidewaystable}[h]
\footnotesize
\scalebox{0.8}{
\begin{tabular}{c|c|cc|cc|cc|cc||c|cc|cc|cc|cc}
\hline \hline
  \multicolumn{9}{c}{CHROMOSOME 21}&  & \multicolumn{8}{c}{CHROMOSOME 22}\\
 \hline \hline
  &   $n$   & A [\%] &  $\sigma_C$ [A] & C[\%] & $\sigma_C$ [C] &  G[\%] & $\sigma_C$ [G] &  T[\%] & $\sigma_C$ [T]  &  $n$   & A [\%] &  $\sigma_C$ [A] & C[\%] & $\sigma_C$ [C] &  G[\%] & $\sigma_C$ [G] &  T[\%] & $\sigma_C$ [T]\\
   \hline
\multirow{5}{*}{$k=1$} &2& 32.52 &  9.24 & 19.66 &  7.38 & 21.05 &  7.01 & 28.31  & 9.95&2& 28.21 &  8.29&  23.37 &  5.53&  24.47&   5.44&  25.65&   9.69\\
 &4 &32.27 & 10.84 & 19.25 &  7.69 & 21.18  & 8.50 & 27.29 & 10.81&4& 28.30  & 8.40&  22.93 &  5.81 & 24.63  & 5.53  &24.92 &  8.48\\
 &6& 32.20 & 10.90 & 18.38  & 8.19 & 20.98  & 8.07 & 28.45 & 10.73& 6&27.59  & 8.41&  23.98 &  6.54&  24.10&   5.87 & 24.81  & 7.87\\
 &8& 32.41 & 10.32 & 17.72  & 7.97 & 21.53  & 8.50 & 28.44  &10.97& 8&26.79 &  7.54  &24.46  & 7.00&  23.85 &  6.70 &25.19  & 7.22\\
 &10& 33.28 &  9.08 & 16.28  & 7.82 &  23.14 &  9.14 & 27.29  & 9.70& 10&26.91  & 7.33 & 24.28  & 7.20 & 23.32 &  7.10 & 25.91 &  7.13\\
   \hline
\multirow{5}{*}{$k=2$} &2& 31.33  &12.48 & 20.18  & 8.33 & 21.33 &  8.27 & 27.94 & 12.05&2& 28.72  & 7.15&  23.14 &  6.23 & 22.19 &  6.52&  26.81&   6.35\\
 &4& 30.79 & 12.03 & 20.27 &  8.92&  21.61 &  8.30 & 27.61 & 12.33& 4&28.05  & 6.28 & 22.47&   5.96 & 22.82  & 6.22 & 26.80&   7.09\\
 &6& 31.16 & 11.19 & 19.70 &  7.91 & 20.83 &  7.86 & 28.42 & 11.55&6& 28.21&   6.46 & 22.49&   7.13&  23.53 &  7.29 & 26.03 &  7.54\\
 &8& 31.73 & 10.24 & 19.58  & 7.46 & 20.48  & 7.80 & 28.21 & 11.21& 8&28.16  & 6.65&  22.20&   7.20 & 23.21&   7.10&  26.44 &  7.23\\
 &10& 31.96 &  9.91 & 19.43  & 7.70 & 20.50 &  7.75 & 28.27 & 10.63& 10& 27.97  & 5.58 & 22.28&   5.32 & 23.39 &  6.14 & 26.35&   6.65\\
   \hline
\multirow{5}{*}{$k=3$} &2 &28.33 & 10.45  &22.50 &  9.69 & 24.70 &  8.74 & 25.18  & 9.31&2& 25.65 &  6.75 & 25.32&   7.10  &23.36  & 6.74 & 25.83&   8.30\\
 &4& 28.38 & 10.78 & 22.33 &  9.23 & 24.33  & 9.51 & 25.69 &  9.84&4& 25.56 &  6.21 & 25.26 &  6.48 & 23.93 &  6.33 & 25.74 &  7.70\\
 &6& 27.95 &  9.06&  21.98 &  8.10 & 24.01 &  8.76 & 26.06 &  9.32&6& 25.55&   6.23&  25.26 &  6.02&  24.39&   6.13 & 24.94 &  6.98\\
 &8& 28.36 &  8.95 & 21.66 &  7.86 & 24.10 &  8.33 & 25.87  & 8.83&8& 24.90  & 5.56 & 25.50 &  5.78 & 24.92 &  6.67 & 24.67 &  6.40\\
 &10& 27.94  & 8.86 & 21.64 &  7.85 & 24.30 &  9.44 & 26.31 &  8.78&10& 24.71&   5.89&  25.73 &  6.63&  25.08 &  6.38&  24.49&   6.03\\
 \hline \hline
  \multicolumn{9}{c}{CHROMOSOME X}&  & \multicolumn{8}{c}{CHROMOSOME Y}\\
 \hline \hline
  &  $n$   & A [\%] &  $\sigma_C$ [A] & C[\%] & $\sigma_C$ [C] &  G[\%] & $\sigma_C$ [G] &  T[\%] & $\sigma_C$ [T]  &  $n$   & A [\%] &  $\sigma_C$ [A] & C[\%] & $\sigma_C$ [C] &  G[\%] & $\sigma_C$ [G] &  T[\%] & $\sigma_C$ [T]\\
   \hline
\multirow{5}{*}{$k=1$} &2& 38.67 & 14.73 & 25.21 & 11.59 & 26.64 & 13.51  &37.66 & 15.49&2& 27.56 &  5.91&  21.89&   5.63 & 23.92 &  6.35 & 27.13 &  7.52\\
 &4& 32.88 & 18.06 & 20.86 & 14.09 & 25.39 & 14.87 & 28.01 & 19.06&4& 27.45  & 6.19&  22.05 &  6.66 & 24.21 &  7.08 & 26.49 &  7.47\\
 &6& 31.16 &  17.92 & 20.53 & 14.69 & 24.35 & 15.84 & 27.09 & 18.84&6& 27.99 &  6.64 & 21.59 &  6.55 & 24.40  & 7.23 & 26.02 &  7.50\\
 &8& 30.65 & 18.29 & 20.27 & 15.59 & 24.21 & 15.68 & 26.05 & 18.37&8& 28.12 &  7.02 & 21.55 &  7.17 & 24.14 &  7.85 & 26.19 &  8.48\\
 &10& 30.12  &17.79 & 20.48 & 14.47 & 22.97 & 15.27 & 27.38 & 18.34&10& 27.89 &  6.99 & 21.88 &  6.60 & 23.30  & 7.52 & 26.90 &  8.10\\
   \hline
\multirow{5}{*}{$k=2$} & 2&41.63 & 15.41 & 20.02 & 10.78 & 25.17  &14.33 & 38.13  &13.19&2& 31.50 &  5.49 & 19.11 &  3.89 & 19.90  & 5.15 & 29.49 &  5.97\\
 &4& 36.55 & 19.80  &19.34&  14.29 & 23.76 & 14.87 & 30.55 & 18.56&4& 31.25  & 5.32 & 19.39  & 4.36 & 19.72  & 4.72 & 29.64 &  5.89\\
 &6& 34.19&  17.37 & 18.30 & 13.05 & 24.36 & 14.39 & 26.35 & 17.62&6& 31.00 &  5.51 & 19.38 &  4.23 & 20.45 &  4.93 & 29.17 &  5.71\\
 &8& 32.40 & 17.61 & 18.84 & 12.59 & 23.18 & 13.59 & 26.49 & 17.30&8& 30.82  & 5.48 & 19.57  & 4.31 & 20.34 &  4.65 & 29.28 &  5.38\\
 &10& 31.63&  15.34 & 18.49 & 11.41 & 23.45 & 13.08 & 26.43 & 15.56&10& 30.83 &  5.16 & 19.28 &  4.26 & 20.68 &  4.24 & 29.20  & 4.85\\
   \hline
\multirow{5}{*}{$k=3$} &2& 40.67 & 14.60 & 23.74 & 14.84 & 22.29 & 13.32&  39.75  &13.89&2& 31.33 &  5.54 & 19.29 &  4.94 & 20.09  & 4.91 & 29.29  & 6.08\\
 &4& 33.73 & 17.43 & 21.57 & 12.97&  21.01 & 13.04 & 32.29 & 15.92&4& 31.61 &  6.17&  18.90 &  4.98 & 20.53  & 5.72&  28.96 &  6.95\\
 &6& 29.39 & 17.92 & 22.04 & 13.80 & 19.14 & 12.31  &31.79 & 16.95&6& 30.91 &  5.64 & 19.35 &  4.42 & 20.25 & 4.41 & 29.49  & 5.88\\
 &8& 29.00 & 17.57 & 21.27 & 12.02 & 19.26 & 11.20 & 32.49 & 15.59&8& 30.76 &  5.34&  19.29 &  3.98 & 20.34  & 4.47 & 29.60 &  6.02\\
 &10& 28.61 & 16.07 & 20.97&  11.35 & 19.76 & 10.58  &31.25 & 15.08&10& 30.61 &  5.37 & 19.19 &  4.15 & 20.46 &  4.59 & 29.73&   5.77\\
 \hline \hline
\end{tabular}}
{\footnotesize  TABLE S6. Same as Table 1 but for the chromosomes 21, 22, 23, 24.}
\end{sidewaystable}

\end{document}